\pdfoutput=1
\documentclass[pdftex]{egpubl}
\usepackage{pg2020}

\ConferenceSubmission   % uncomment for Conference submission
\SpecialIssueSubmission    % uncomment for submission to , special issue
\usepackage[T1]{fontenc}
\usepackage{dfadobe}
\usepackage{amsmath}
\usepackage{amssymb}
\usepackage[boxed]{algorithm}
\usepackage[noend]{algpseudocode}
\usepackage{booktabs,dcolumn,caption}
\usepackage{multirow}
\usepackage{verbatim}
\usepackage{subcaption}
\usepackage[export]{adjustbox}

\usepackage{cite}  % comment out for biblatex with backend=biber
\BibtexOrBiblatex
\electronicVersion
\PrintedOrElectronic
\ifpdf \usepackage[pdftex]{graphicx} \pdfcompresslevel=9
\else \usepackage[dvips]{graphicx} \fi

\usepackage{egweblnk}
\title[SDFDDGI]%
      {Signed Distance Fields Dynamic Diffuse Global Illumination}

\author[submit 3253061]
{\parbox{\textwidth}{\centering Jinkai Hu \qquad Milo Yip \qquad G. Elias Alonso \\ Shihao Gu \qquad Xiangjun Tang \qquad Xiaogang Jin
        }
{\parbox{\textwidth}{\centering AAA BBB CCC
       }
}
}

\begin{document}
 \teaser{
    \vspace{-15px}
    \centering

    \begin{subfigure}{0.324\textwidth}
    \includegraphics[width=1\linewidth]{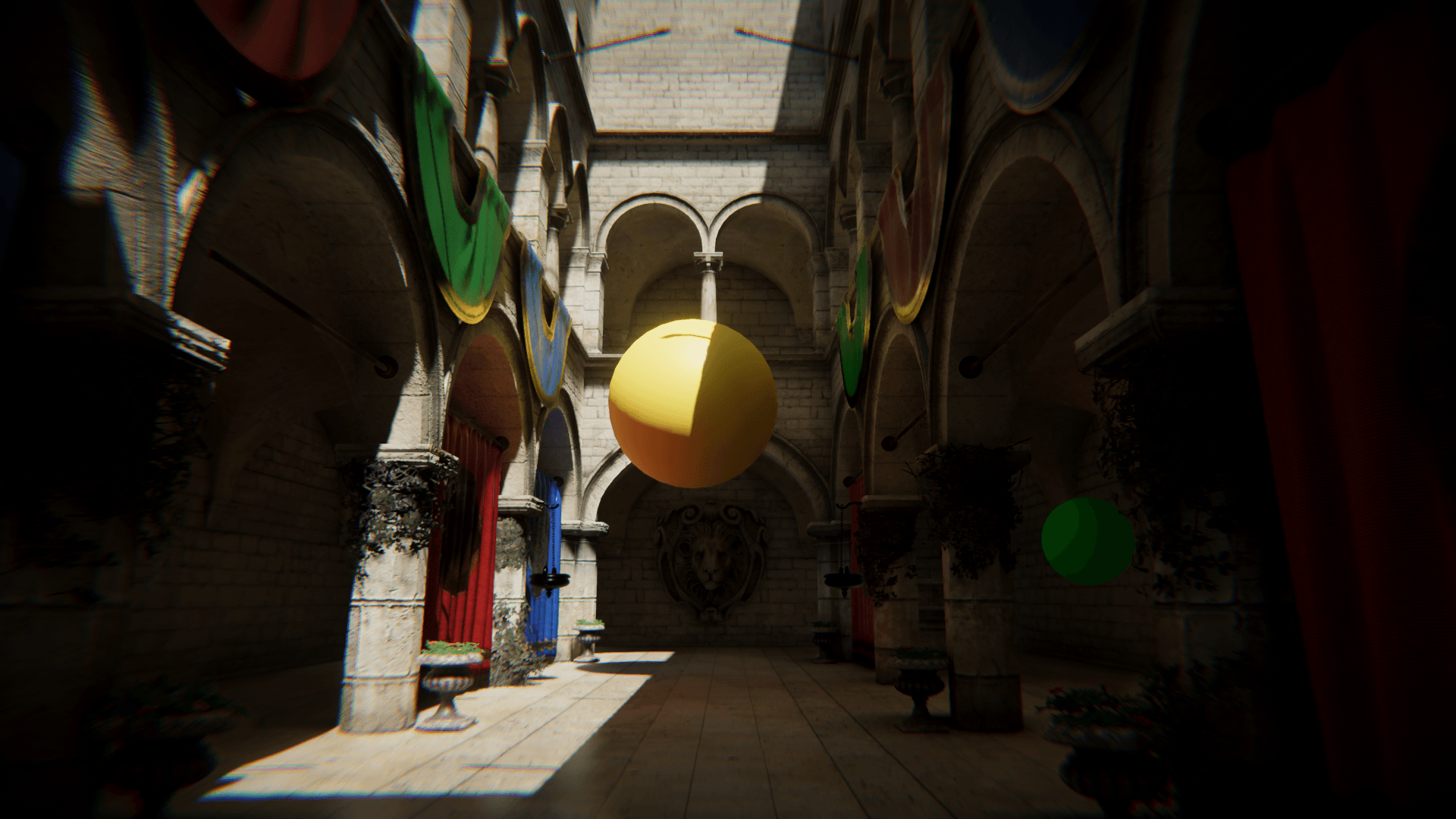}
    \caption{SDFDDGI full result}
    \end{subfigure}
    \begin{subfigure}{0.324\textwidth}
    \includegraphics[width=1\linewidth]{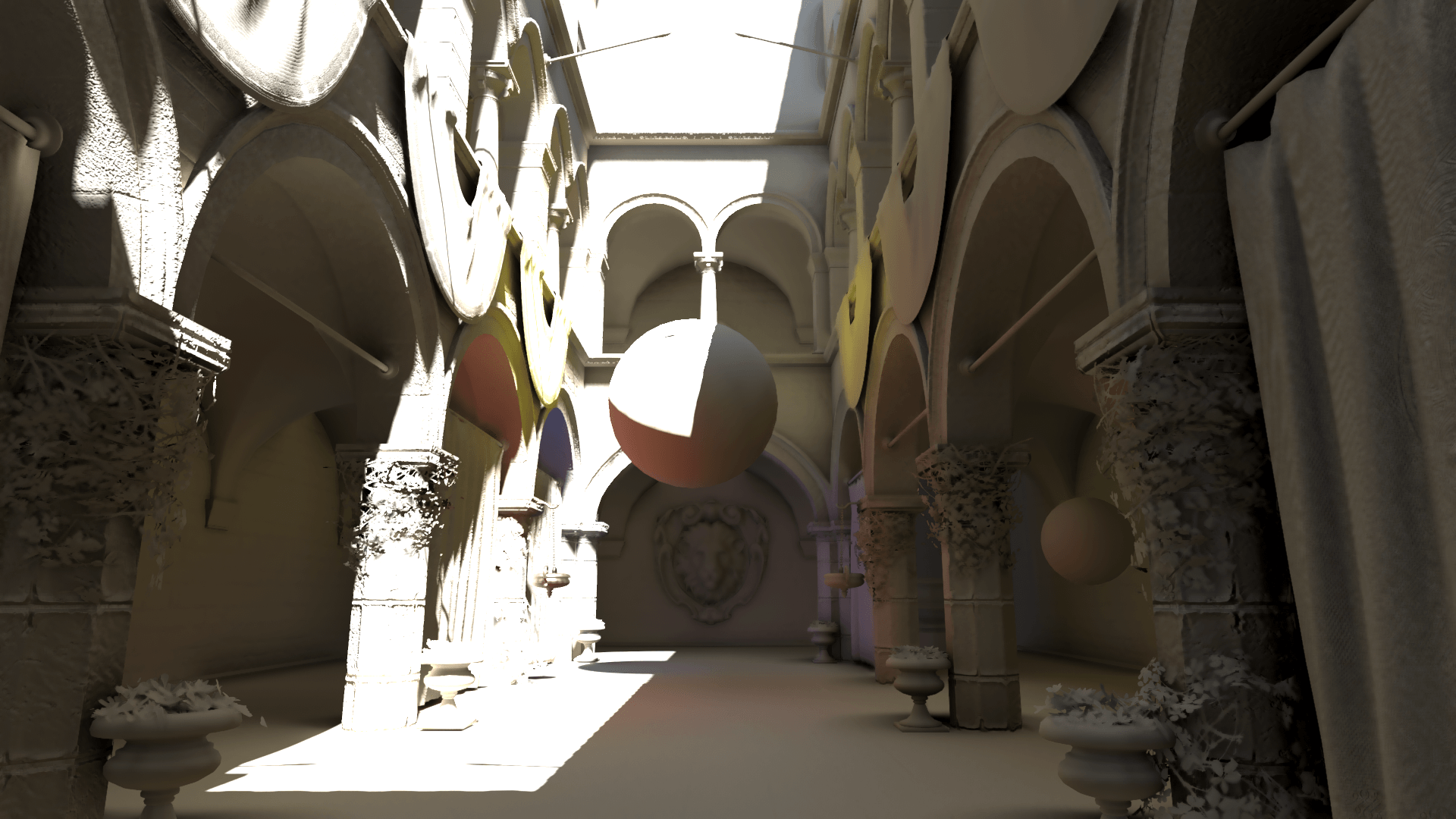}
    \caption{SDFDDGI illumination}
    \end{subfigure}

    \begin{subfigure}{0.16\textwidth}
    \includegraphics[width=1\linewidth]{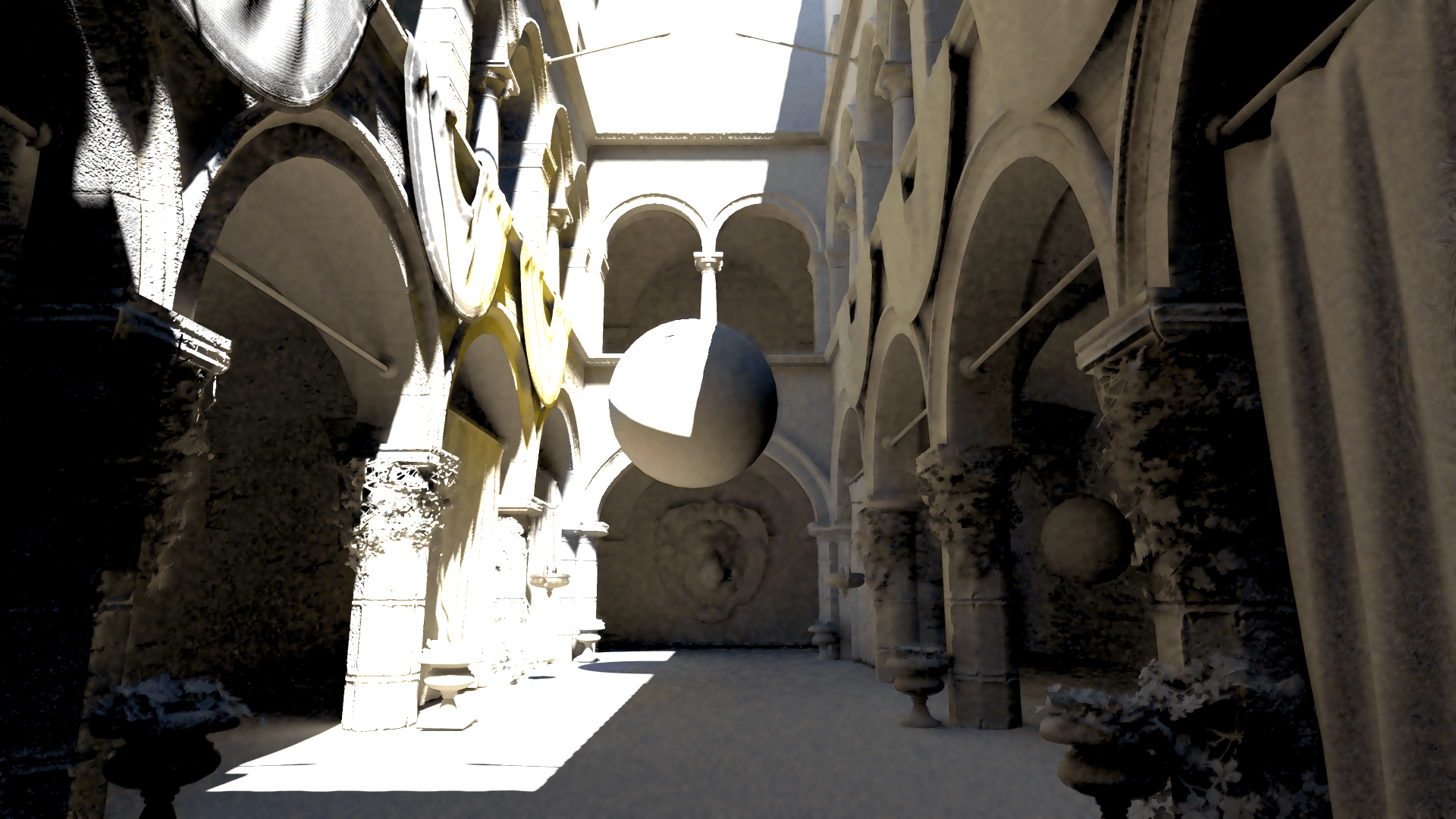}
    \caption{Ray Tracing GI}
    \end{subfigure}
    \begin{subfigure}{0.16\textwidth}
    \includegraphics[width=1\linewidth]{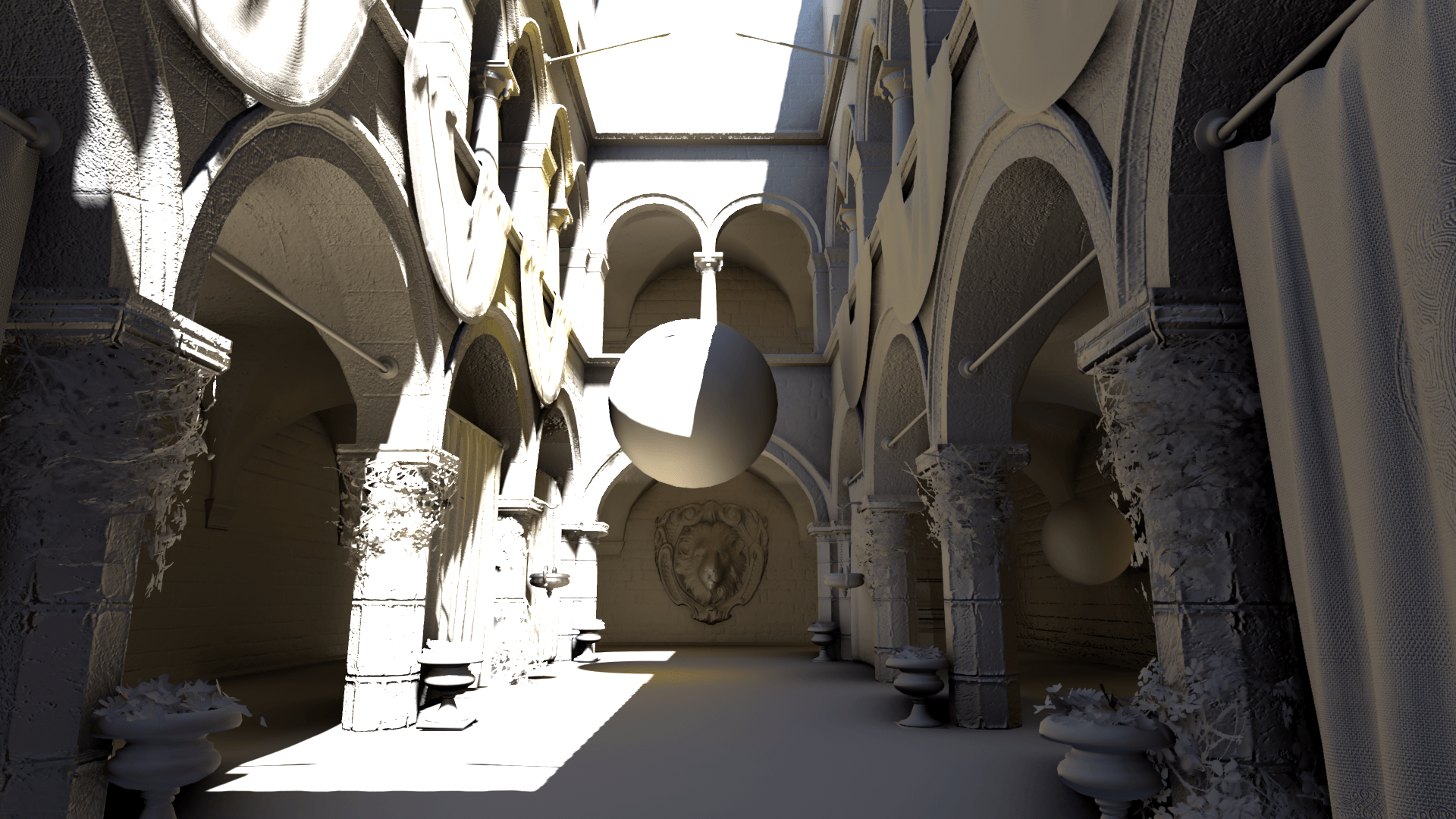}
    \caption{RTXGI}
    \end{subfigure}
    \begin{subfigure}{0.16\textwidth}
    \includegraphics[width=1\linewidth]{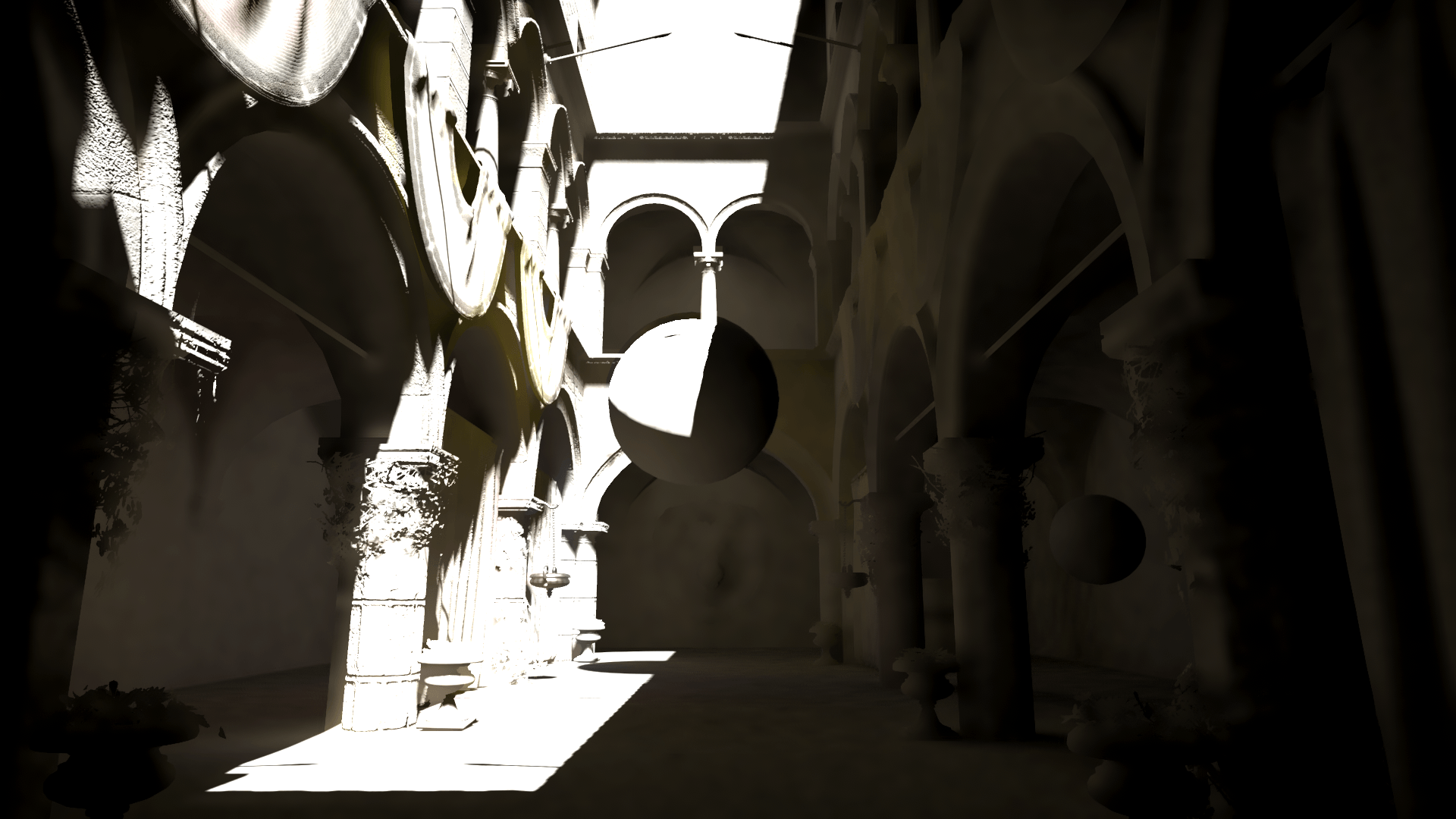}
    \caption{SSGI}
    \end{subfigure}
    \begin{subfigure}{0.16\textwidth}
    \includegraphics[width=1\linewidth]{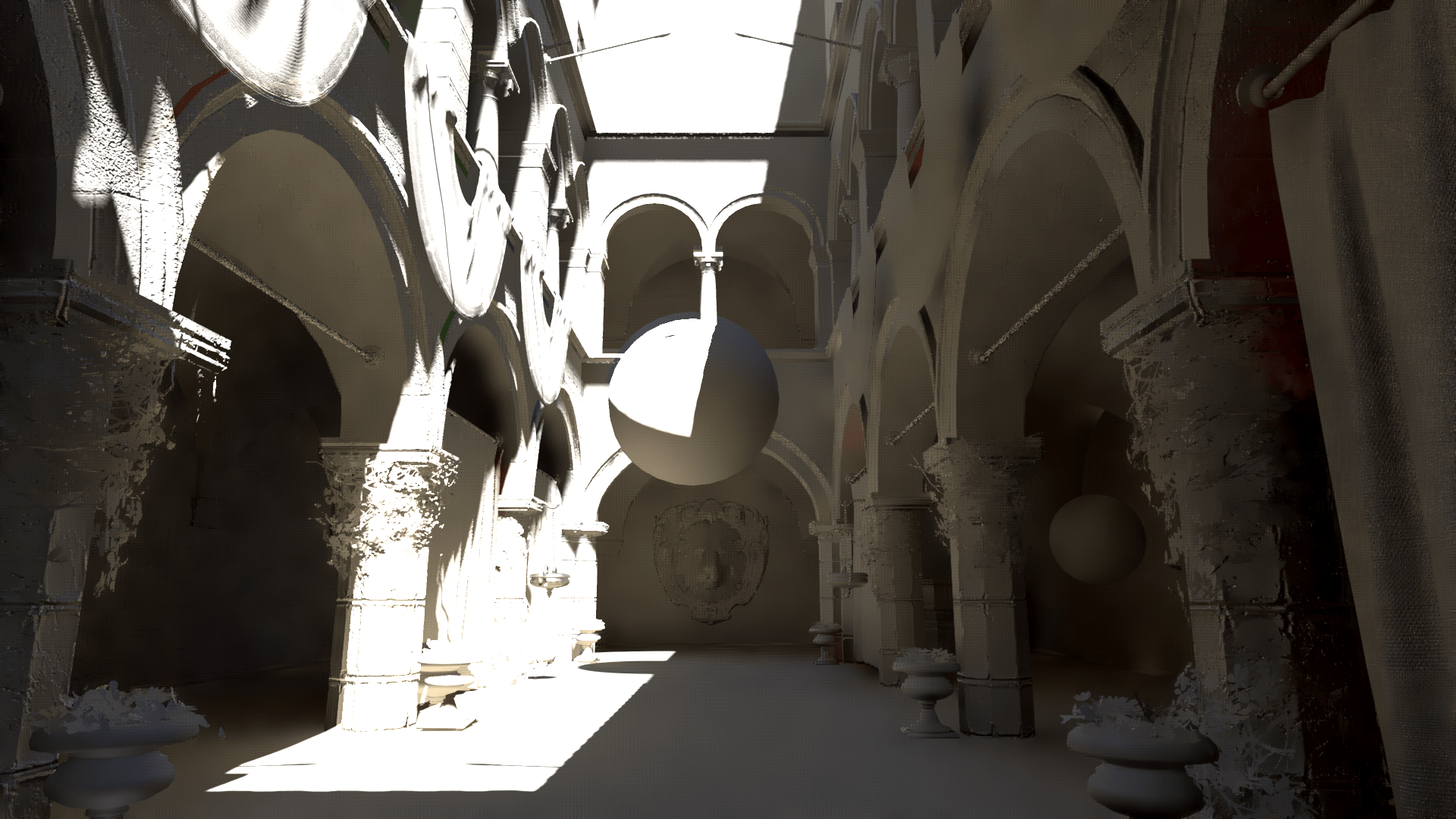}
    \caption{VXGI}
    \end{subfigure}

    \centering
    \caption{The figure above provides an example of our SDFDDGI on the upper row (a)(b) as well as a comparison to other real-time GI methods in the row below. Our approach achieved less than 5 ms per frame on GTX 970M hardware with the lowest acceptable quality, while on RTX 2080Ti even achieved a performance within 1 ms per frame. Besides, it solves other method's deficiencies such as real-time Ray Tracing's extra noise and lack of multi-bounce lighting (a), light leaking issues on the dynamic yellow sphere on RTXGI (b), mere use of screen space information in Screen Space GI (c), and rougher detailing on Voxel-Based GI (d).}
    %我们方法和其他方法的结果对比。左上：Full result. 右上：Lighting Only. (a) Ray Tracing Diffuse GI, 缺少多次反射，且有噪点影响. (b) RTXGI, 运动的球体上有漏光发生，且细节不佳. (c) Screen Space GI, 仅有屏幕空间的GI反射信息，效果不佳. (d) Voxel-Based GI, 细节质量不佳，有噪声影响。
    \label{fig:teaser}
}

\maketitle

%-------------------------------------------------------------------------
\begin{abstract}
   Global Illumination (GI) is of utmost importance in the field of photo-realistic rendering. However, its computation has always been very complex, especially diffuse GI. State of the art real-time GI methods have limitations of different nature, such as light leaking, performance issues, special hardware requirements, noise corruption, bounce number limitations, among others. To overcome these limitations, we propose a novel approach of computing dynamic diffuse GI with a signed distance fields approximation of the scene and discretizing the space domain of the irradiance function. With this approach, we are able to estimate real-time diffuse GI for dynamic lighting and geometry, without any precomputations and supporting multi-bounce GI, providing good quality lighting and high performance at the same time. Our algorithm is also able to achieve better scalability, and manage both large open scenes and indoor high-detailed scenes without being corrupted by noise.

% CCS codes.


\end{abstract}
%-------------------------------------------------------------------------
\section{Introduction}

Global illumination allows us to improve strikingly the realism of a virtual scene. However, its real-time computational cost is far too expensive for most applications, such as in the video game industry.

To solve this problem, many approaches have been proposed. For example, baked Light Maps\cite{mctaggart2004half} allow us to compute global illumination for static lights and scenes by baking the lighting of the scene. Precomputed Radiance Transfer\cite{sloan2002precomputed} solved their limitations for dynamic light sources. \cite{seyb20uberbake} proposed an extension for the baking system of Light Maps that supports dynamic changes. However, all these approaches pose limitations on the dynamic changes, thus affecting the artistic design of the scenes. Besides, precomputations add an extra production cost.

Of course, there also exist some methods that can manage completely dynamic GI, for example: Voxel-Based GI\cite{crassin2011interactive}, Ray Tracing GI\cite{oles2019exploring}, Screen Space GI\cite{ritschel2009approximating}\cite{silvennoinen2015multi}\cite{sousa2011secrets}, RTXGI\cite{majercik2019dynamic} and so on. These techniques fill the gap left by the area of real-time global illumination, making it possible to use GI for real-time applications, like in the video game industry. However, these methods have limitations of different nature, such as light leaking, performance issues, special hardware requirements, noise corruption, bounce number limitations and lack of scene information, among others.

Inspired by RTXGI\cite{majercik2019dynamic}, we propose SDFDDGI (Signed-Distance-Field Dynamic Diffuse Global Illumination). Compared to RTXGI, our approach doesn’t need the use of specific hardware, made improvements in its performance and result, increased dynamic response speed and completely solved light leaking problems.

This method uses Signed Distance Fields (SDF)\cite{Hart1996sphereTracing} to make a slim representation of the scene. We sample the irradiance function of the space domain and interpolate these samples to estimate its global illumination. To improve the resemblance of the original irradiance function and its discrete form, we employ SDF to lead the distribution of sampling points. Then, when interpolating samples, we use SDF for the visibility tests to prevent the ocurrence of light leaking artifacts. We also use Contact GI to enhance details on SDFDDGI.

Our paper makes the following contributions:

\begin{itemize}
\item We use SDF primitives and clusters to compress the information in the scene, reduce data size, increase memory use and speed up computations.
\item SDF allows us to query very fast the nearest distance and spatial gradient to optimize the position of discretized sampling points in order to better suit the spatial distribution of the irradiance function.
\item SDF is able to compute soft shadows efficiently\cite{aaltonen2018gpu}, optimize the interpolation strategy of the discretized points, and completely avoid light leaking artifacts, as well as provide soft indirect shadowing effects.
\end{itemize}

%-------------------------------------------------------------------------
\section{Related work}

Since the appearance of the rendering equation\cite{kajiya1986rendering}, global illumination has historically shaped the whole development of graphics theory.

\begin{equation}
L_o(P, D_o)=L_e(P,D_o)+\int_{\Omega}f_r(P,D_i,D_o)L_i(P,D_i)(n\cdot D_i)\mathrm{d}D_i.
\end{equation}

In the field of real-time photorealistic rendering, the speed and quality of the computation of global illumination has all along been a hot topic of research. In this section, we present a general overview of some of the most concerning GI techniques and research.

Virtual Point Lights (VPL)\cite{laine2007incremental} is one of the earliest real-time GI technique. Its main contribution is the substitution of GI computations by adding virtual point light sources at the areas illuminated by direct lighting. This approach has many limitations but its result is satisfactory for local indirect illumination of spotlight light sources or other narrow-ranged ones.

Reflective Shadow Maps (RSM)\cite{dachsbacher2005association} is also an early method for real-time global illumination. It is based on the same idea as shadow maps, not only store depth information but also store light sources direction and radiant flux as well and use this reflective shadow map as global illumination. However, it does not take into account indirect occlusion, thus producing severe indirect illumination misestimations in some scenes. It also only provides one-bounce GI and is not able to manage area lights and skylight.

RSM and VPL ultimately evolved into Light Propagation Volumes (LPV)\cite{kaplanyan2010cascaded}. LPV introduced the concept of volume in VPL and transfers illumination data across space. This method solved many of the problems of VPL and RSM, but light leaking remained a serious concern as well as some accuracy issues.

Apart from this, there are voxelization approaches\cite{crassin2011interactive}, which first voxelize the scene into a sparse voxel tree and then inject lighting data. This allows us to estimate global illumination at real-time frame rates but it also produces light leaking. Besides, in scenes with highly varying dynamic geometry, the computational cost of voxelization is too high.

High-end global illumination approaches like Ray Tracing GI\cite{oles2019exploring} reconstruct world coordinates and normals out of G-buffers, then sample the hemisphere to compute global illumination. The number of computations of this kind of approaches is very large, so it is only viable for a small amount of samples and high-end hardware, thus needing a final denoising stage\cite{schied2017spatiotemporal}\cite{koskela2019blockwise} for an acceptable result. The main disadvantage of this method is its performance, being almost impossible to calculate multi-bounce illumination, apart from the cost of additional denoising.

At the moment of writing this, performance-wise the best choice is Screen Space GI, like screen space diffuse GI\cite{ritschel2009approximating}\cite{silvennoinen2015multi}, as well as screen space specular reflections (SSR)\cite{sousa2011secrets}. Comparitively speaking, necessary information for specular reflections normally resides inside screen space, while diffuse GI often lacks most needed lighting information, thus not being able to provide an optimal result.

Last year, NVIDIA proposed a new approach RTXGI\cite{majercik2019dynamic} using its ray tracing accelerated hardware, by means of discretizing the spatial distribution of the irradiance function. Compared to common probe-based GI\cite{mcauley2015rendering}, its main contribution is the use of depth information and Variance Shadow Maps (VSM)\cite{donnelly2006variance} as well, in order to prevent light leaking artifacts that arise from the discretization of irradiance. However, its effect on GI of details at real-time frame rates is not optimal. Besides, light leaking artifacts can also appear with very thin objects and it depends severely on RTX-accelerated hardware, which affects its use extent.

On that basis, we realised that SDF\cite{Hart1996sphereTracing} can be used to simplify the scene representation for low-frequency global illumination like diffuse GI. SDF is a scalar field in the space domain, which represents the distance from a point in space to the nearest surface in the scene. A positive value is assigned if the point is in the outer region of the nearest surface and negative if it is inside, thus producing a compact representation of the geometry information of a scene.

Inspired by RTXGI, we proposed a novel approach SDFDDGI, which overcomes aforementioned limitations and has the following advantages:

\begin{itemize}
\item It does not need any precomputations.
\item It can manage both dynamic geometry and dynamic lighting, as well as animations and skylight.
\item It provides interframe stability and low delay response for dynamic changes.
\item It completely eradicates light leaking problems.
\item Our technique is not limited to specific hardware, it can also be used in lower-end hardware.
\end{itemize}

%-------------------------------------------------------------------------
\section{Approach}

SDFDDGI has mainly 4 stages:

\begin{enumerate}
\item \textbf{SDF Cluster}: generate the scene's SDF representation and its clustering acceleration structure.
\item \textbf{Probe choosing}: choose suitable sampling points in space.
\item \textbf{Probe update}: update the irradiance on a sphere of directions around the sampling points.
\item \textbf{Per pixel GI shading}: interpolate different probes to calculate global illumination for all pixels in the screen.
\end{enumerate}

%-------------------------------------------------------------------------
\subsection{SDF Cluster}
SDF is usually stored in volume textures. However, with this approach we need to make a trade-off between the resolution and detail of the scene and the data size. Therefore, we describe our scene with a distance field composed of different SDF primitives, then parse the global SDF value in space instead of performing the voxelization of the scene. For example, we only need 4 KB to store an SDF representation of the Sponza Palace scene.

Our scene objects are represented by a series of basic SDF primitives. Each SDF primitive includes a primitive type and its respective transform. Example primitive types are rectangular blocks, planes, cylinders, or any other whose analytical form is simple enough to easily compute its SDF value. In every frame, we use CPU to perform culling on SDF primitives on the surroundings of the camera and perform Level Of Detail (LOD) for further away primitives, i.e. use less primitives to express a larger rougher shape. Afterwards, we generate a cluster structure to speed up the SDF query.

\begin{figure}[h]
    \centering
    \includegraphics[width=1\linewidth]{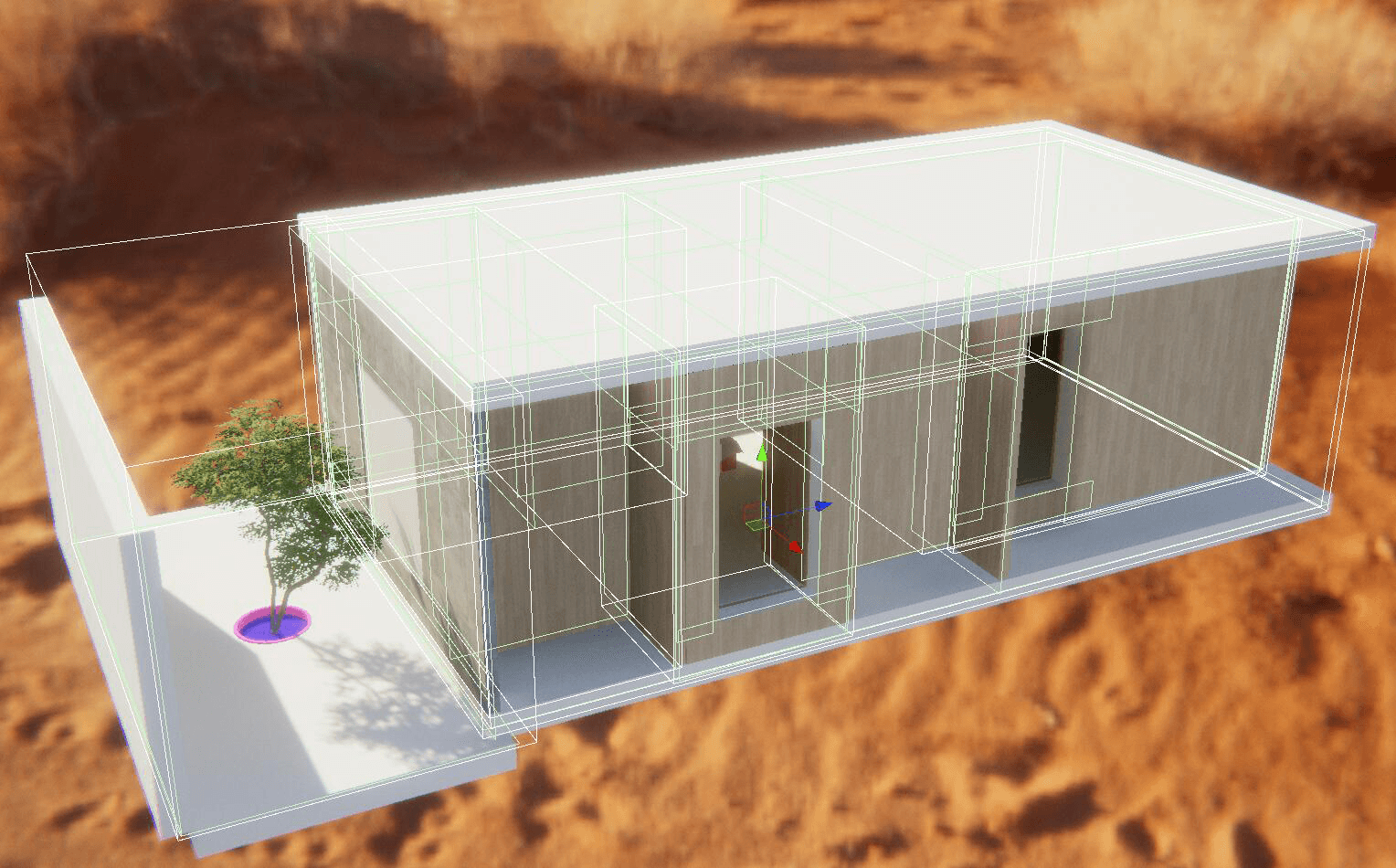}

    \caption{For the computation of the diffuse global illumination, the scene is represented as a set of different SDF primitives, marked as green lines in the example picture. Then, near SDF primitives are packed into a cluster, marked as white lines above, to accelerate the query of the scene SDF value at one point.}
    \label{fig:sdfandcluster}
\end{figure}

We use a clustering approach to pack near SDF primitives into a cluster, as shown is Figure \ref{fig:sdfandcluster}. When performing a query, SDF primitives in a cluster will first be treated as one and be jointly rejected or not, what refrains us from querying every single SDF primitive, thus accelerating the process by 20\% to 100\% depending on the scene. This has proven to have a better performance than BVH \cite{wald2007ray} for smaller scale data. Since the number of primitives is not much, we use the CPU to better distribute the strain of the GPU. Besides its time expenditure is negligible in comparison to the other stages.

%我们使用聚类方法，将距离相近的SDF primitives打包成Cluster，并在query时对Cluster进行快速剔除，这样的加速结构在数据规模小时，相对于常用的BVH结构\cite{wald2007ray}有更好的性能. 通过使用Cluster进行剔除，相比于暴力For Loop，在不同场景下我们获得了20%~100%的性能提升。

In today’s hardware, memory access cost is far greater than the cost of computations. Unordered access of memory, which is characteristic of diffuse GI, has an even more negative impact on performance on cache. In ray tracing algorithms we often need to use Ray Binning\cite{anis2019leveraging} and other techniques to improve cache hit ratio. However, in our case the primitives are relatively small, so much as to being able to put everything inside the L1 cache, making SDF query even more advantageous than reading volume textures. Moreover, this implies that we can easily support dynamic scenes, not like voxelization approaches that

At the same time, since the data we need is mainly on the L1 cache, global memory is available for use for other parts that have lower requirements on memory access. In this manner, processes that have high requirements on memory but are computationally inexpensive, like G-Buffer or Shadow Map generation, can run on the GPU, thus increasing GPU general utilization rate, which is an important factor on the general frame renderization time.

%-------------------------------------------------------------------------
\subsection{Probe choosing}

Since Irradiance is a $\mathbb{R}^5$ function, we split its domain in two parts: the space coordinates $\mathbb{R}^3$ and the direction $\mathbb{R}^2$ as Equation \ref{con:ID}.

\begin{equation}
\begin{split}
P =\begin{bmatrix} x \\ y \\ z \end{bmatrix}, D =\begin{bmatrix} \omega \\ \theta \end{bmatrix}, \\
Irradiance = E(P, D).
\label{con:ID}
\end{split}
\end{equation}

Our method discretizes the spatial domain of the irradiance function \emph{E} into many sampling points \emph{P}. Every sampling point is referred to as probe. Every probe stores the irradiance on a sphere of directions around its position. We create a spatially arranged probe volume around the camera and use these probes to interpolate global illumination.

Irradiance is not a continuous function in its spatial domain. For example, in walls or at occluded points there are evident discontinuities, which are the main cause of light leaking in most real-time global illumination algorithms. If we want an appropriate representation of the irradiance distribution in space, we need to carefully place the probes.

Because of this reason, we first calculate the SDF value at probe’s location. If it is smaller than a threshold value, this means it is too near to other objects or even inside one, what would negatively impact the quality of the sampling. Then, we query the gradient of SDF and use gradient descent method to obtain an acceptable sampling point near the original position. If the displacement between the position of last and current frame surpasses a threshold, the irradiance at last frame will need to be rejected, so we allocate more rays to this probe to ensure a more stable result.
\begin{figure}[h]
    \centering
    \begin{algorithm}[H]
    \begin{algorithmic}[1]

    \Procedure{UpdateProbePos}{}
    \State $\textit{lastPos} \gets \text{position of }\textit{probe}\text{ in last frame}$
    \State $\textit{pos} \gets \text{position of }\textit{probe}\text{ in the uniform grid}$

    \If {$\text{querySdf}(pos) < \textit{threshold1}$}
    \State $\textit{pos} \gets \text{gradientDescent}(SDF, pos)$
    \EndIf

    \If {$\text{distance}(pos, lastPos) > \textit{threshold2}$}
    \State $\text{markRejectHistory}(probe)$
    \EndIf

    \State \Return $\textit{pos}$
    \EndProcedure
    \end{algorithmic}
    \end{algorithm}
    \caption{Pseudocode for the algorithm to find a suitable position for a probe. We move our probe from the grid if it is too near or inside objects and we check if the displacement of the probe from last frame's position is too large to consider GI information of last frame.}
    \label{fig:updateProbeAlg}
\end{figure}
In order to decrease the number of probe updates, in each frame we first choose which probes need to be updated and divide the probe updates between different frames so as to increase performance.

Different weights are assigned to the probes according to their distance to the camera and direction to the camera in order to decide which probes should be updated.

Due to the probe updates, this method may cause jitter between frames. Nevertheless, this phenomenon is not as evident as in the original RTXGI since the performance of our approach allows us to perform a more extensive sampling, what partly mitigates the effect of jittering.

In contrast to RTXGI, our method does not need human intervention in order to have a better probe distribution and it can rapidly and accurately respond to scene changes. This also reduces the leaking caused by dynamic objects.

%-------------------------------------------------------------------------
\subsection{Probe update}
To obtain the irradiance function \emph{E}, radiance \emph{L} needs to be computed first. We use compute shaders to sample the radiance over a sphere of directions at each probe:
\begin{equation}
E(P,D)=\int_{4\pi}\max(0,\cos(D, D'))\cdot L(P,D') \mathrm{d}D'.
\label{con:ICR}
\end{equation}

In this phase, we use a 8x8 thread block to update each chosen probe. Before sampling, all threads in a block need to cooperate to move all or part of the SDF primitives into the L1 cache, since they will be queried frequently.

\begin{figure}[h]
    \centering
    \begin{subfigure}{0.49\linewidth}
    \includegraphics[width=1\linewidth]{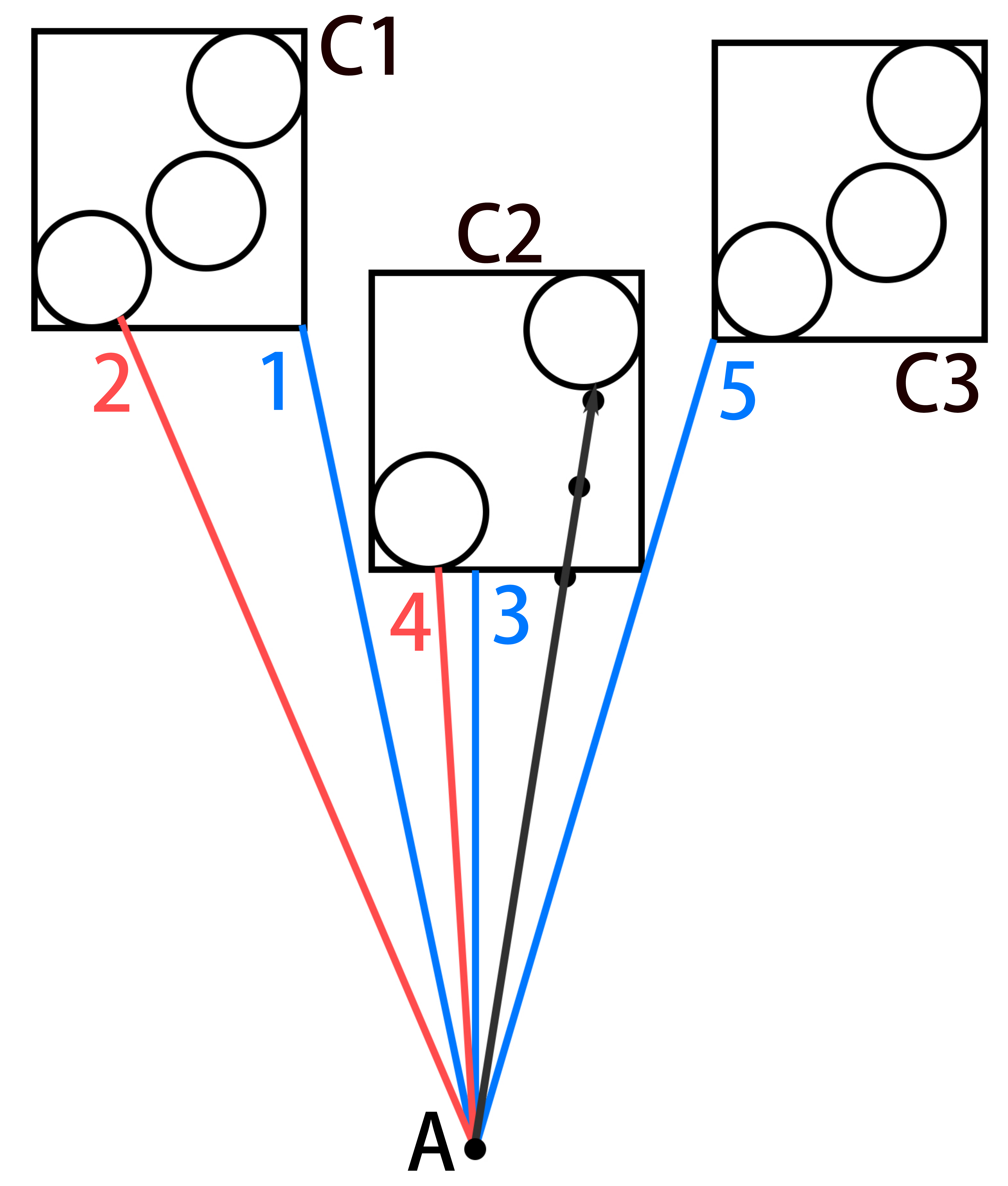}
    \caption{SDF query at point \textit{A} }
    \end{subfigure}
    \begin{subfigure}{0.49\linewidth}
    \includegraphics[width=1\linewidth]{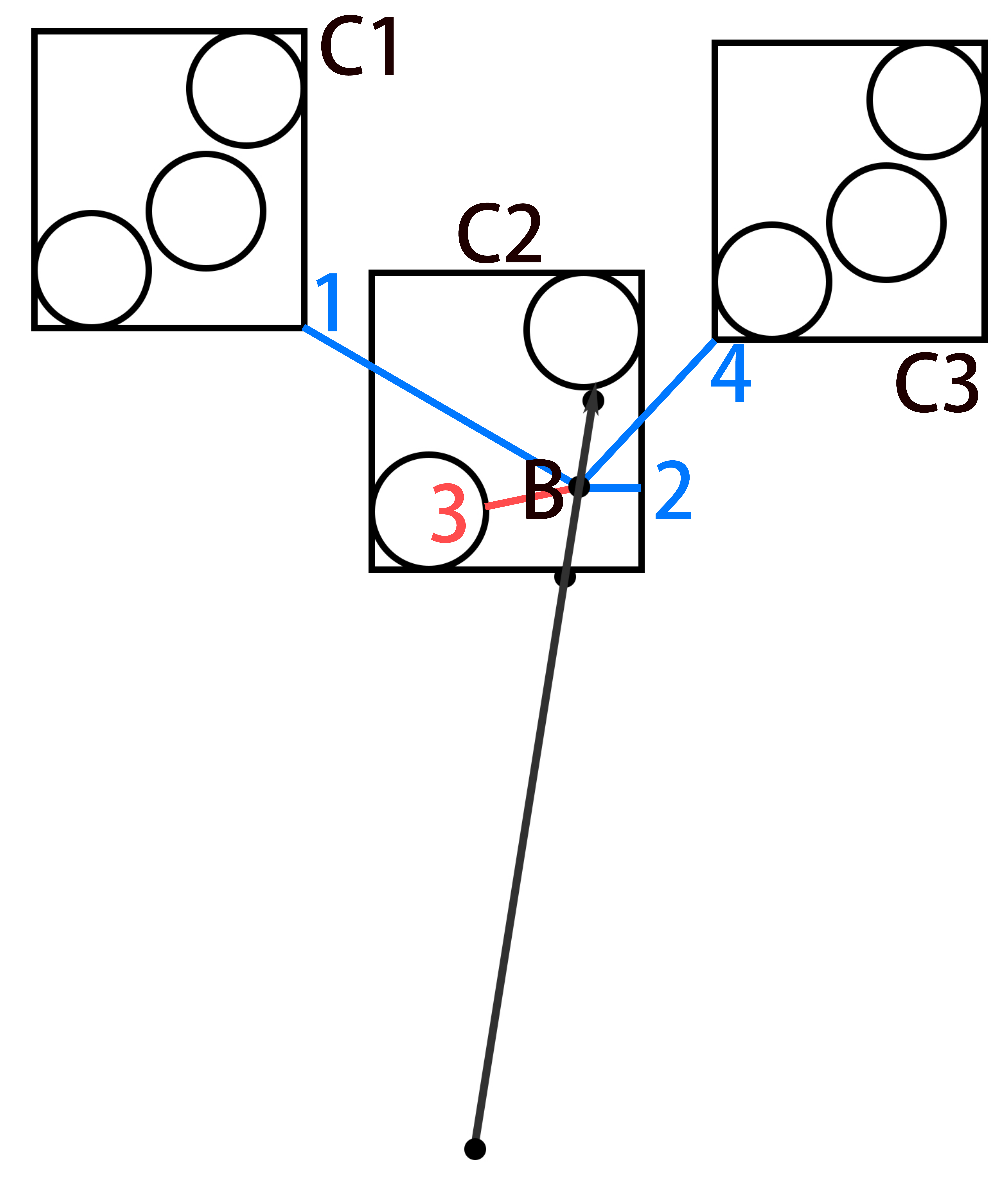}
    \caption{SDF query at point \textit{B} }
    \end{subfigure}

    \caption{This figure provides a 2D example of SDF querying for two consecutive points \textit{A} and \textit{B} for a scene with three clusters \textit{C1}, \textit{C2} and \textit{C3} with their respective SDF primitives inside. The blue lines represent the smaller distance to the bounding box of each cluster while the red lines represent the smaller distance to each cluster's SDF primitive. The numbers enumerate each of the steps of the algorithm.}
    \label{fig:clusters}
\end{figure}

In order to speed up the tracing process, we employ clusters to perform culling on SDF primitives. Taking point \textit{A} in Figure \ref{fig:clusters} as the starting point of tracing (a), the algorithm to query SDF can be described as follows:

\begin{enumerate}
    \item Compute the smallest distance \textit{d1} to the bounding box of the first cluster \textit{C1}.
    \item Since \textit{d1} is smaller than initial distance \textit{D} (initialized here as infinity), traverse all primitives in cluster \textit{C1}, find the nearest distance \textit{d2} to them and update distance \textit{D=min(D, d2)}.
    \item Calculate the smallest distance \textit{d3} to cluster \textit{C2}.
    \item Since \textit{d3} is smaller than \textit{D}, traverse all primitives inside cluster \textit{C2}, find the nearest distance \textit{d4} to them and update distance \textit{D=min(D, d4)}.
    \item Since the distance to the bounding box of cluster \textit{C3} is larger than \textit{D}, we skip this cluster, so SDF query result is \textit{D=d4}.
\end{enumerate}

This algorithm targets a single SDF query but it can be employed together for a series of queries in a ray for further taking advantage of this culling process. Take point \textit{B} as example (b), we can initialize our minimum distance \textit{D} to twice the previous SDF value \textit{D=2lastSDF}, since the SDF value of point \textit{B} needs to be smaller than twice the previous SDF. So the process of querying the SDF at point \textit{B} can be described as follows:

\begin{enumerate}
    \item Get the smallest distance \textit{d1} to cluster \textit{C1}, and since it is greater than \textit{D=2·lastSDF}, skip it.
    \item Compute the smallest distance \textit{d2} to the bounding box of cluster \textit{C2}.
    \item Since distance \textit{d2} is smaller than 0, get the smallest distance \textit{d3} to the primitives inside the cluster and update distance \textit{D}, as it is smaller.
    \item Then, since distance \textit{d4} to cluster \textit{C3} is greater than \textit{D}, we also skip this cluster, so SDF query result is \textit{D=d3}.
\end{enumerate}

As aforementioned, many clusters can be rejected using twice the previous SDF query result as initial distance, especially when SDF value is smaller. This works notably good for sphere tracing, where query density is very high near object boundaries, i.e. where SDF value is small. Thus, algorithm on Figure \ref{fig:clusters} is able to greatly boost the performance of SDF sphere tracing, especially for complex scenes. Besides, this algorithm is stack free and GPU-friendly.

We sample the irradiances at random low-discrepancy\cite{keller2013quasi} directions and store them in the shared memory. To calculate the intersections with the scene, we perform an accelerated version of sphere tracing\cite{korndorfer2014enhanced}\cite{balint2018accelerating} for our circumstances.

After intersection point is calculated,  Reflective Shadow Map (RSM) is used to obtain its flux. Additionally, the emission value stored inside the primitive is used in order to support area lights and self emission.

%为了加速Tracer过程，我们使用Cluster方法对SDF Primitives进行剔除。假设A为Trace的起点。在A点时，SDF查询的算法流程如下：1. 对Cluster1的包围盒计算最近距离d1。2. d1小于初始猜测D=Inf，则遍历Cluster1内所有primitives求出最近距离d2，并更新D=min(D, d2)。3. 对Cluster2的包围盒求出最近距离d3。4. d3小于D=d2，则遍历Cluster1内所有primitives求出最近距离d4，并更新D=min(D, d4)。5. 对Cluster3的包围盒求出最近距离d5,d5大于D=d4，跳过，则最近距离为D=d4。这一算法针对单次SDF Query进行了速度的提升，但注意到对于Ray上的连续的一系列点，该算法仍有进一步优化的空间。

%以B点为例，初始猜测D将等于前一个查询点的SDF值的两倍，因为B点的SDF值必定小于B点前一个点的SDF值的两倍。则B点的SDF查询流程如下：1. 对Cluster1的包围盒求出最近距离d1，d1大于初始猜测D=2*lastSDF，跳过。2. 对Cluster2的包围盒计算最近距离d2。3. d2小于0，则遍历Cluster2内所有primitives求最近距离d3，并更新D=min(D, d3)。5. 对Cluster3的包围盒求出最近距离d4,d4大于D=d3，跳过，则最近距离为D=d3。
%可以看到，通过使用两倍的前一次SDF查询结果作为初始猜测，能够剔除掉大量的Cluster，尤其是在SDF值小的区域。而Sphere Trace过程恰恰在SDF值小时query更密集，这导致了该算法能够极大程度的提升SDF Sphere Trace算法的效率，尤其是在较为复杂的场景下。且该算法是stack free的，对GPU非常友好。

%图 figs/Cluster/AB.png

By reusing probe's GI data of last frame, we can achieve multi-bounce global illumination. Since multi-bounce diffuse GI is generally lower frequency than first bounce diffuse GI, we limit its sample number to a lower level in order to decrease computational cost. Apart from this, we can speed up GI response by multiplying multi-bounce GI by a coefficient between 0 and 1 and, thus, decrease loop gain effects.

After synchronizing all threads in a block, each thread calculates the irradiance in their own direction \emph{D} using Equation \ref{con:ICR}, and use octahedral mapping\cite{cigolle2014survey} to write irradiance data into a probe atlas texture. We keep a balance between the sample amount and the degree of temporal mixing. If there are too few samples, we apply a stronger temporal mixing in order to prevent jitter. By storing the radiance samples to the group shared memory, we can reuse them
across threads to compute the irradiance, thus stabilizing the result.

%-------------------------------------------------------------------------
\subsection{Per pixel GI shading}
Probe Visibility tests take an important role to prevent light leaking triggered by the discretization of irradiance in spatial domain.

RTXGI used probe depth buffers and VSM\cite{donnelly2006variance} to perform these tests, which is limited to the resolution of the depth buffer, thus unable to completely remove the light leaking effect, especially the leaking caused by thin objects. Inspired by the easy generation of soft shadows using SDF\cite{aaltonen2018gpu}, we use the SDF shadow trace to make visibility test, which is able to naturally produce soft indirect shadows and transitions.

Each pixel needs to interpolate 8 probes, what implies to carry sphere tracing 8 times, making it unacceptable as it is too expensive to compute. Fortunately, we only need to perform probe visibility test instead of sampling radiance using sphere trace. For this reason, we can do an extensive down sampling at this stage.

First we down sample the screen depth buffer according to a min/max checkerboard\cite{fabian2019creating} to obtain a half resolution buffer. For checkerboard black pixels, we obtain the maximum value in an area of 4 pixels in the full resolution depth buffer, otherwise for white pixels we take the minimum. This is made to assure we will have valid samples that can cover the whole depth range. This algorithm was proposed in 2019 by the Red Dead Redemption team and it has proven to be very effective for downsampling in the domain of low frequency rendering.

After that, in the min/max checkerboard down sampled depth buffer, out of every 2x2 pixel block we choose one to perform probe visibility test. This choosing process follows the next principles:

\begin{itemize}
\item For different frames, choose a different pixel.
\item Assure as much as possible that we cover the whole depth extent.
\end{itemize}

Before performing visibility test on the chosen pixel, we first perform a duplicate removal on each 2x2 block unit. It is obvious that close starting points have the same visibility test result for a specific probe. Therefore, in every 2x2 block it is necessary to assure that the rest of the probes are not repeated or the distance between starting points is not too large for the visibility test. We use shared memory for the pixels to exchange information, then after performing duplicate removal 4 threads are assigned on average to each pixel inside the block, making every thread’s visibility task diminish from 8 to 4 on average. In our experiments, we use a 4x4 size block to obtain an optimal accelerated result, because an oversized block can decrease the merging of visibility test tasks, thus decreasing the effectiveness of the weighting algorithm.

After completing the visibility test tasks, we write the result back into shared memory and distribute it to every corresponding pixel in order to employ this for probe interpolation. Since our approach is able to accurately obtain the visibility of the probe, we do not need to perform an extra cosine weighting or any other additional weighting terms like RTXGI to further avoid leaking artifacts.

\begin{figure}[h]
    \centering
    \includegraphics[width=1\linewidth]{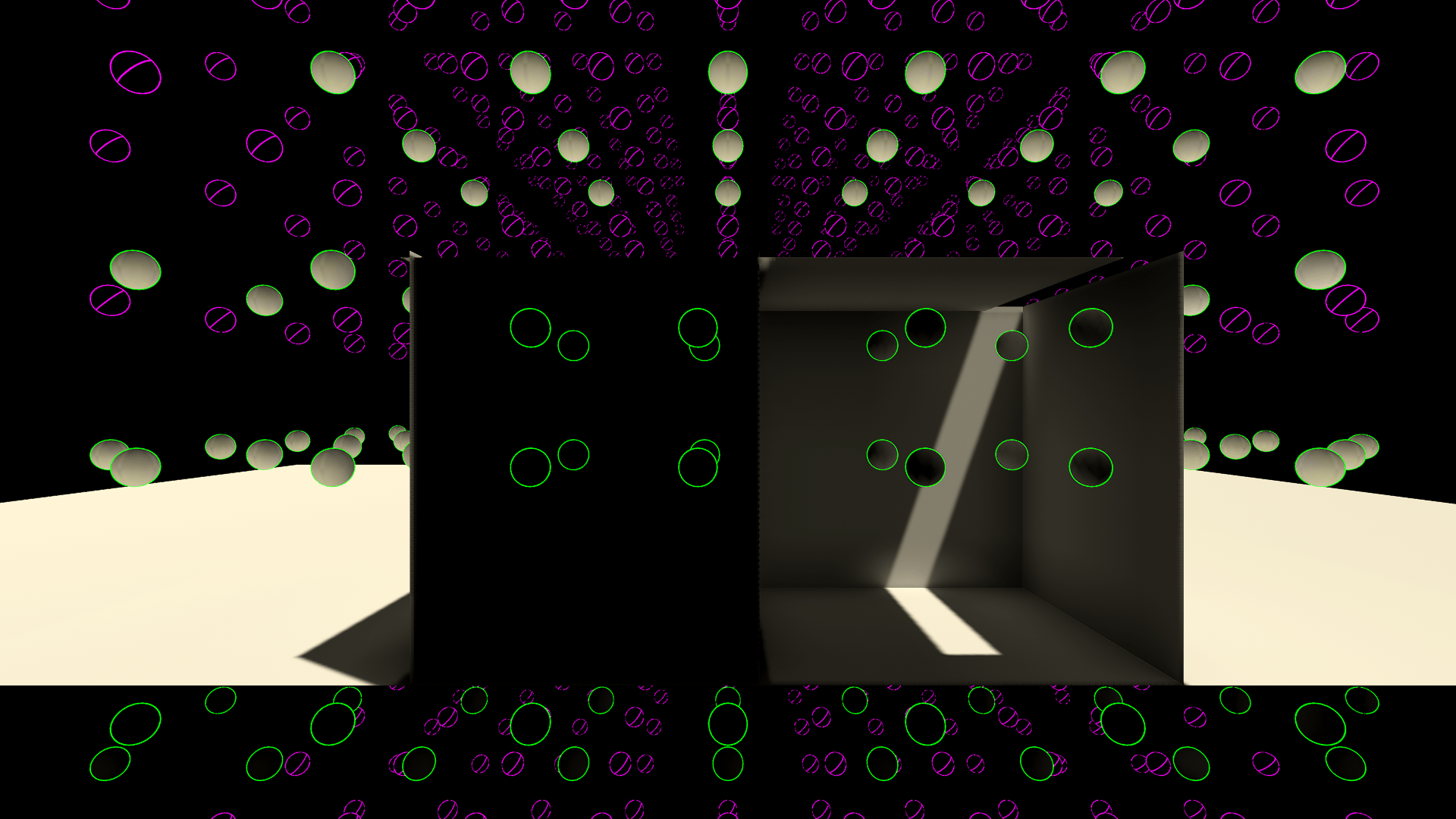}
    \caption{Dark room without light leaking even with thin walls. Our approach is independent of voxel resolution, so objects of any thickness are not able to produce light leaking.}
    \label{fig:leakprevent}
\end{figure}

\begin{figure*}·
    \centering
    \begin{subfigure}{0.49\linewidth}
        \includegraphics[width=1\linewidth]{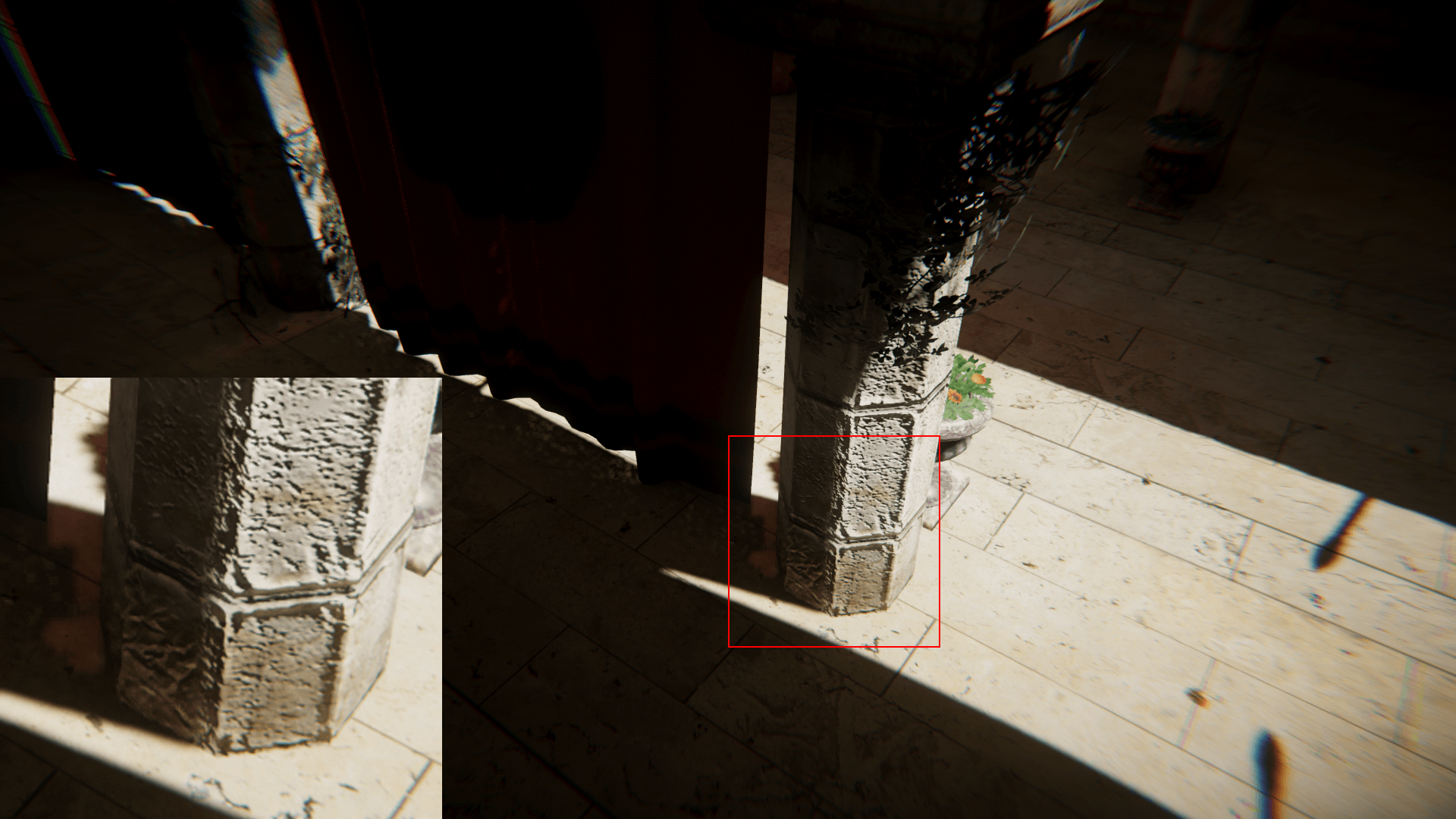}
        \caption{SDFDDGI}
    \end{subfigure}
    \begin{subfigure}{0.49\linewidth}
        \includegraphics[width=1\linewidth]{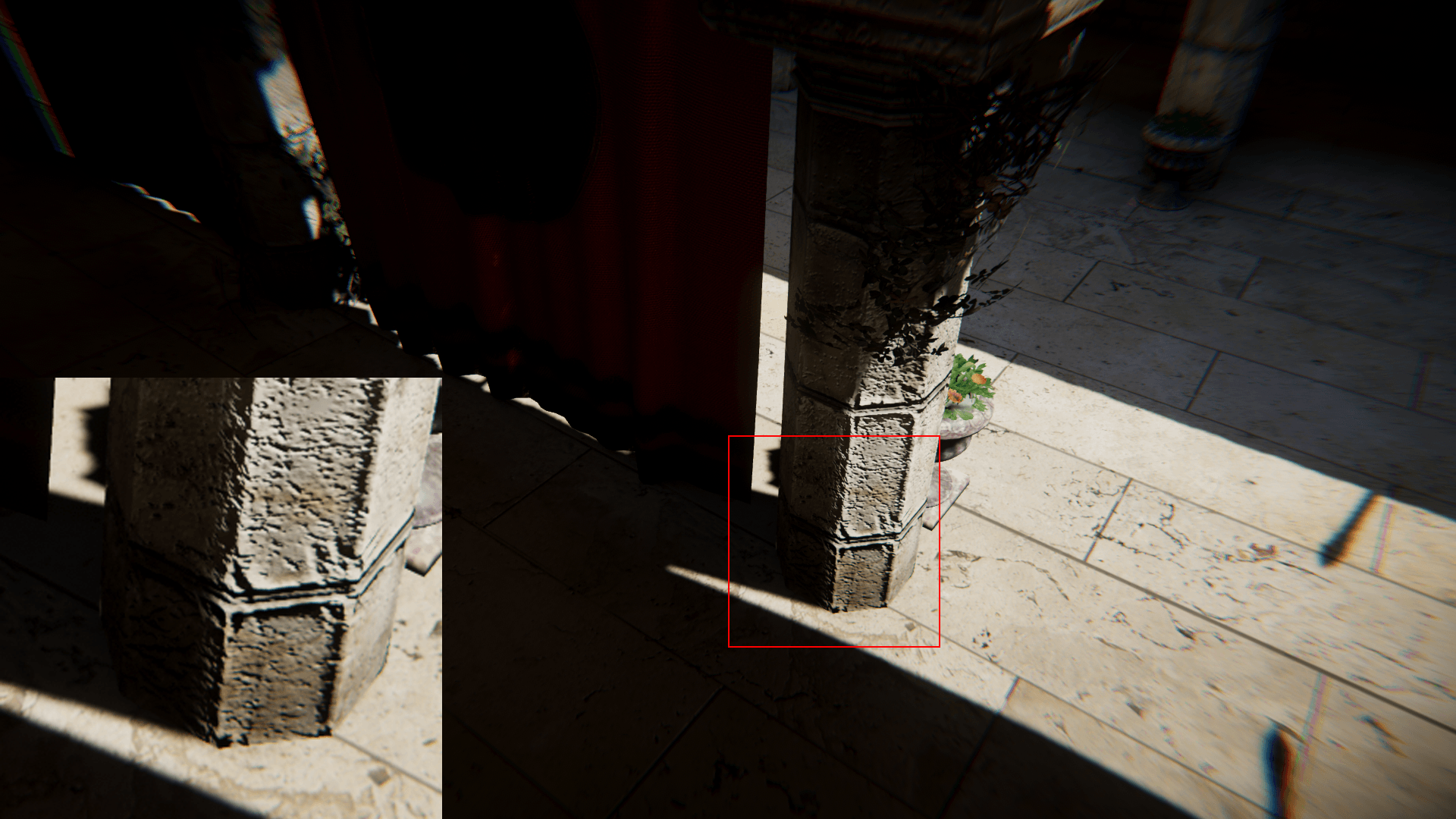}
        \caption{RTXGI}
    \end{subfigure}

    \begin{subfigure}{0.49\linewidth}
        \includegraphics[width=1\linewidth]{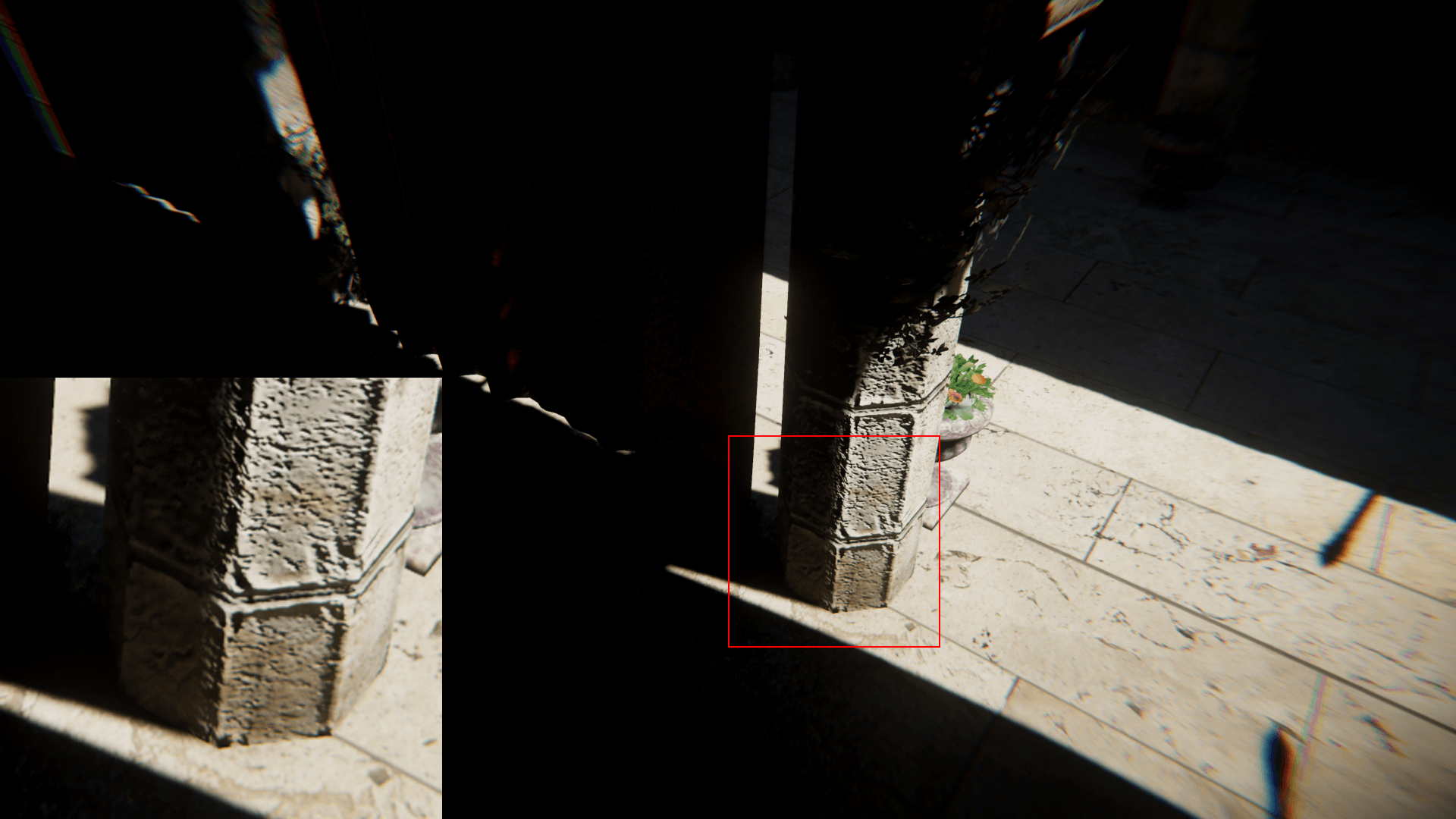}
        \caption{Ray Tracing GI}
    \end{subfigure}
    \begin{subfigure}{0.49\linewidth}
        \includegraphics[width=1\linewidth]{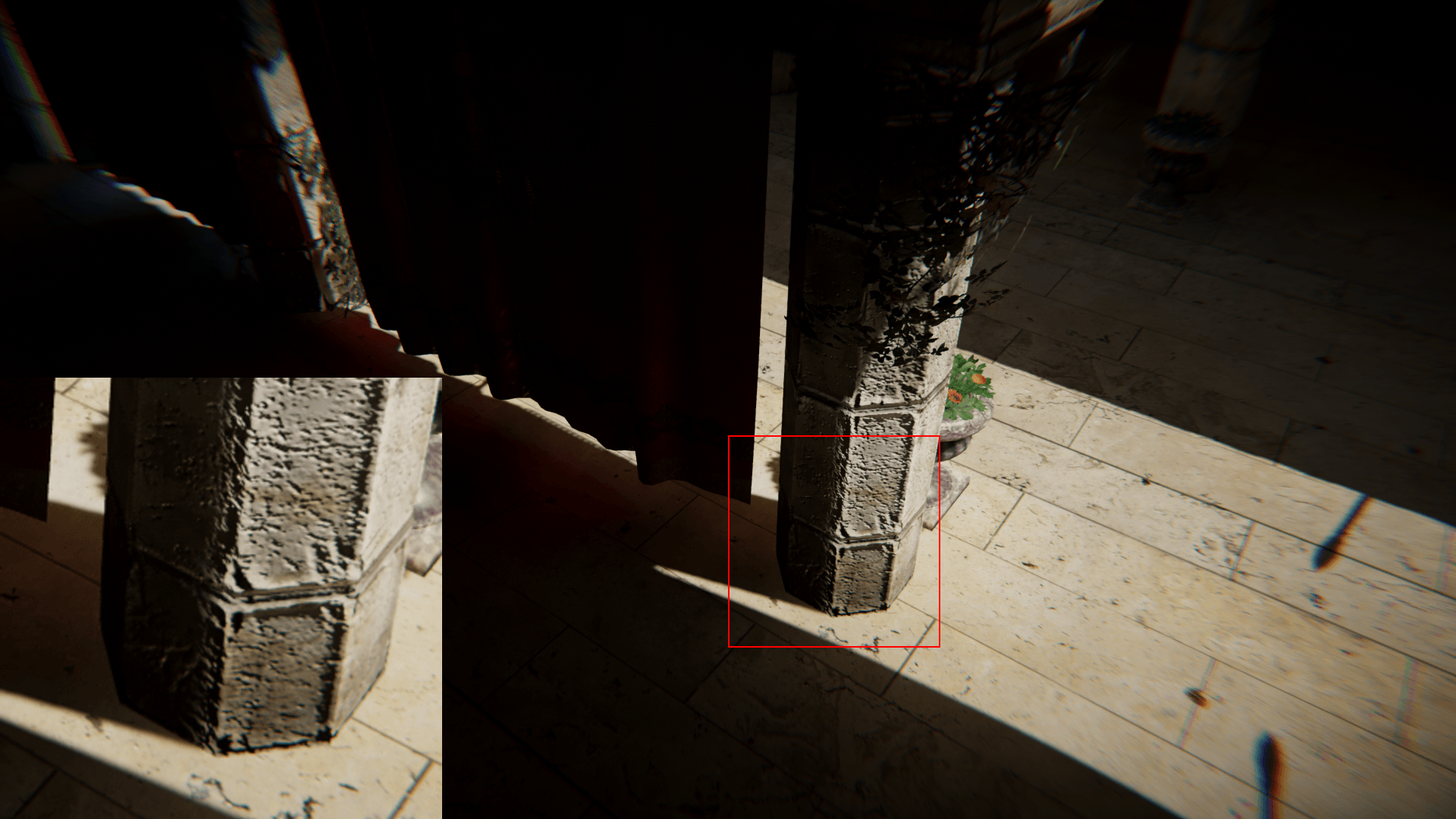}
        \caption{VXGI}
    \end{subfigure}

    \caption{Our approach (a) is able to retain the reflections on the pillar, like for RTXGI (b) and VXGI (d). Ray Tracing GI (c), however, lacks multi-bounce reflections and needs extra denoising, which may blur details.}
    %对比(b)(d),我们的方法保留了柱子处的光线反射，对比(c),我们的方法无需降噪因而能够完整保留柱子法线的细节
    \label{fig:gicompare}
\end{figure*}

In the upsampling stage, if a pixel does not have a valid result, we assign it a value according to surrounding pixels and last frame’s reprojection result using motion vector. What affects more in the whole process to the final quality of the result is finding representative pixels in the second phase, but we can reuse past results and Neighbourhood Clipping\cite{karis2014high} to reduce its influence until it is negligible.

This method greatly compensates the overwhelming strain produced by the per pixel visibility tests. This technique reduces 8 tests per pixel to 0.25 tests per pixel or lower, greatly increasing performance and almost without a negative impact on the end result.

RTXGI uses a low resolution depth buffer to perform these visibility tests, whereas we employ a per pixel probe visibility test to completely eradicate light leaking as Figure \ref{fig:leakprevent} shows and optimizing the process with an effective down sampling to extra reduce the computational cost.

%-------------------------------------------------------------------------
\subsection{Contact GI}

We can greatly diminish the computational cost of GI by discretizing the irradiance on the space domain. However, this leads to loss of detail. We created a new technique called Contact GI based on Ground Truth Ambient Occlusion (GTAO)\cite{jimenez2016practical} to enhance the details of diffuse GI.

Ambient occlusion represents how much ambient lighting receives an object surface according to the occlusion received by surrounding objects. In the per pixel GI shading stage, the position of probes and that of the actual pixel is not the same triggers the loss of global illumination detail. By adding AO to the GI shading result we can to some extent enhance its details. However, this practice does not take into account the effect of multi-bounce indirect illumination among other additional problems. For example, in directly lighted regions the effect of diffuse GI over AO should be greater, but in RTXGI this wrongly weakens the indirect illumination. By using contact GI we corrected this problem.

\begin{figure}[h]
    \centering
    \begin{subfigure}{0.49\linewidth}
        \includegraphics[width=1\linewidth]{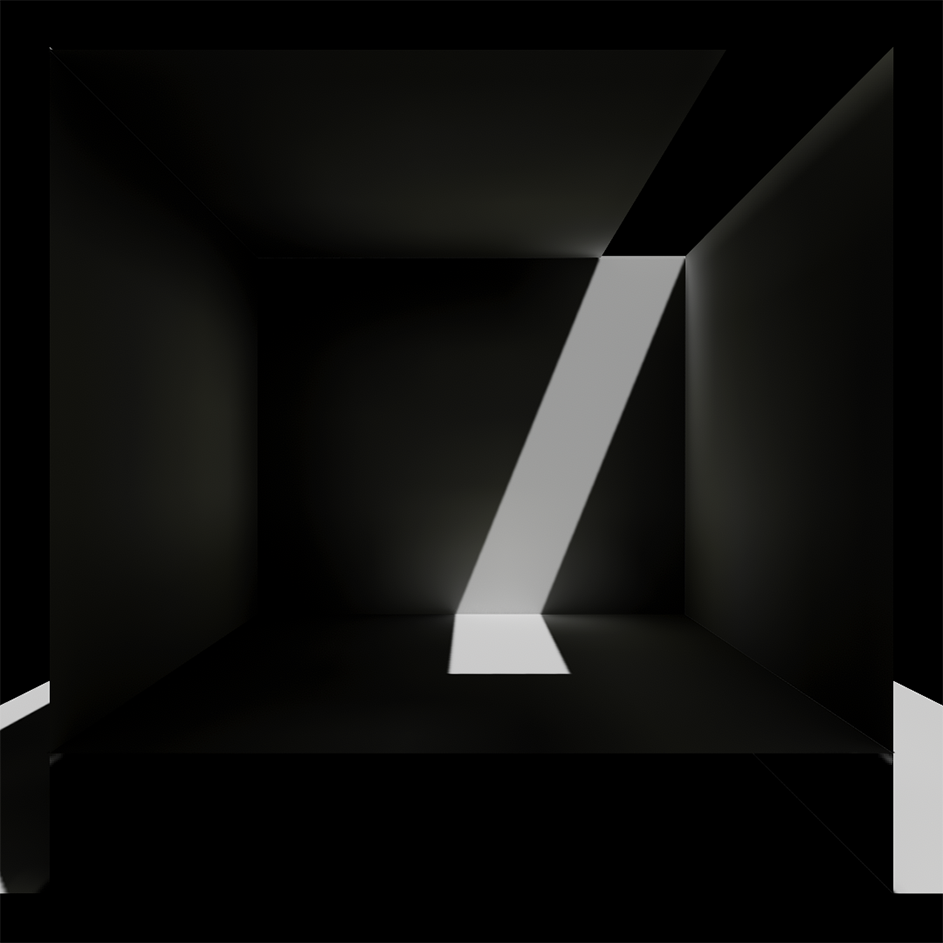}
        \caption{Contact GI on}
    \end{subfigure}
    \begin{subfigure}{0.49\linewidth}
        \includegraphics[width=1\linewidth]{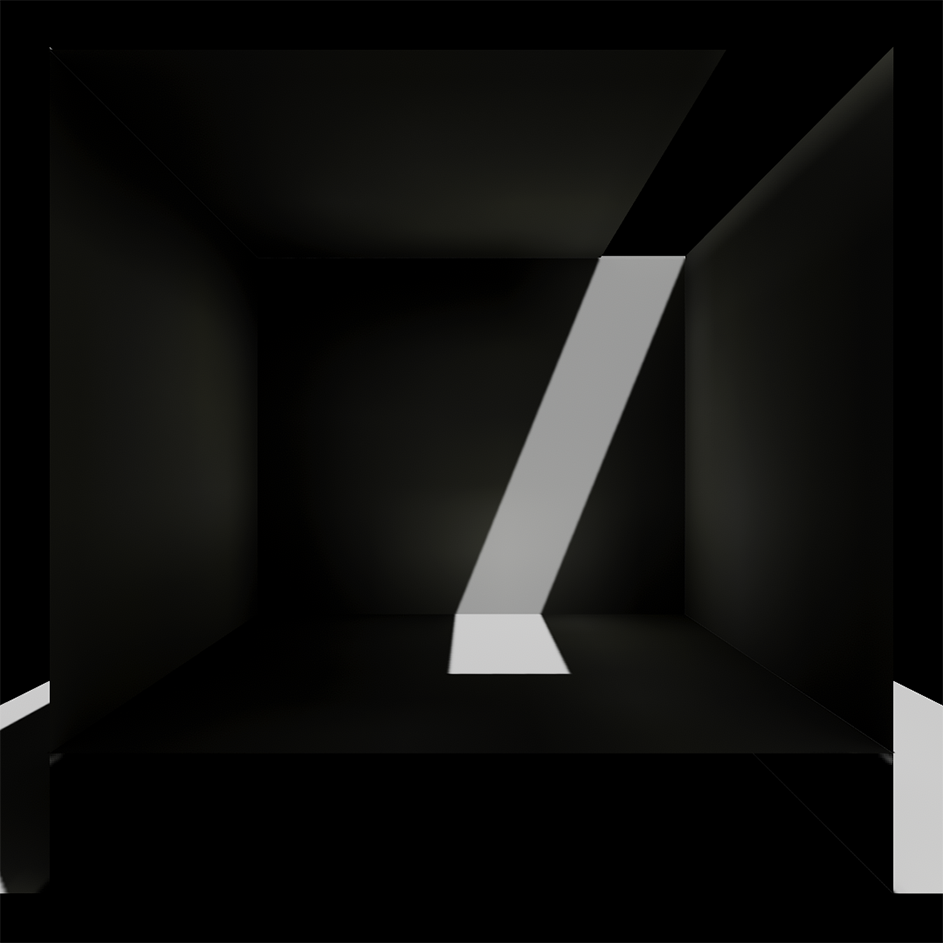}
        \caption{Contact GI off}
    \end{subfigure}

    \begin{subfigure}{0.98\linewidth}
        \includegraphics[width=1\linewidth]{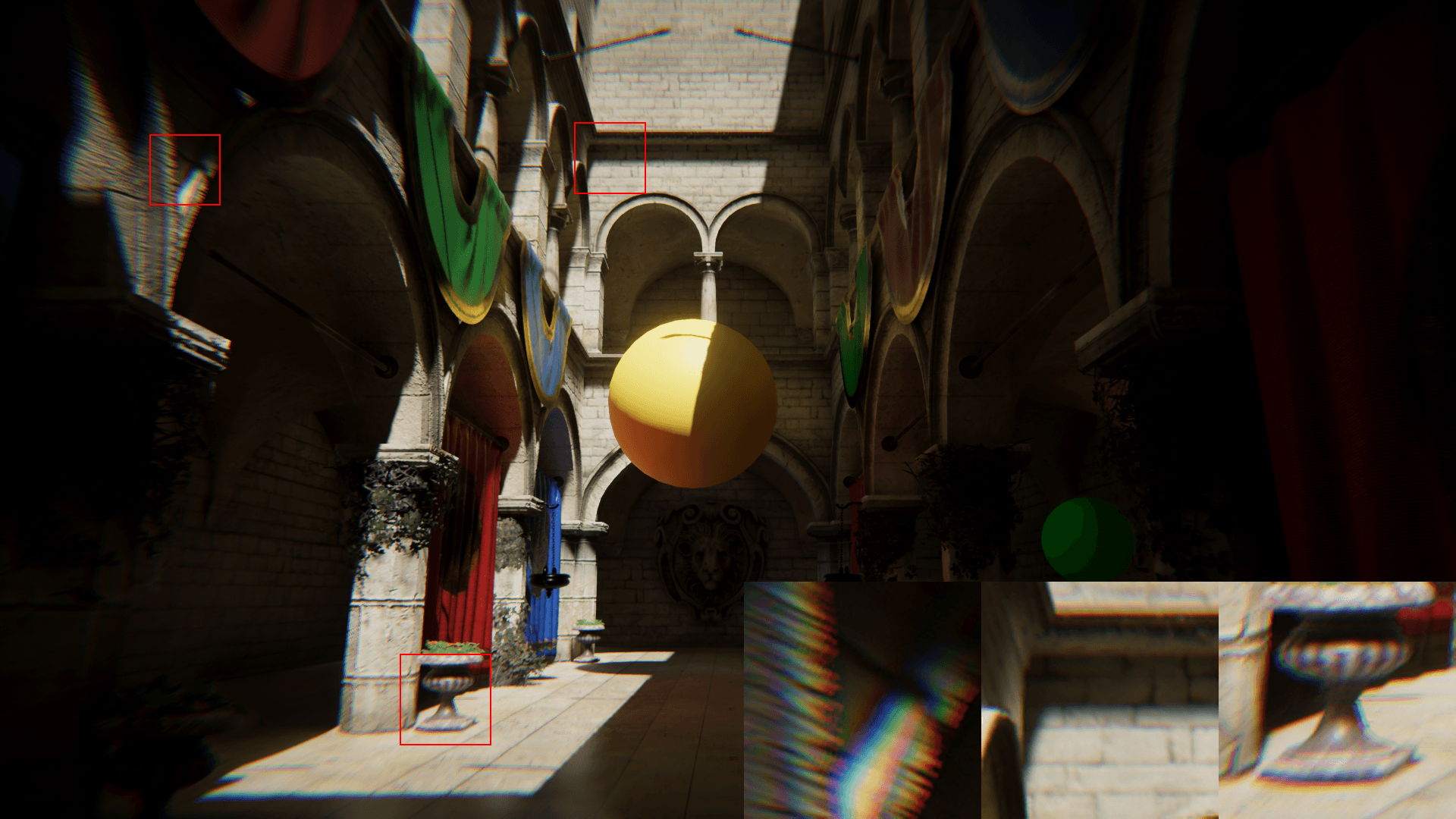}
        \caption{Contact GI on}
    \end{subfigure}

    \begin{subfigure}{0.98\linewidth}
        \includegraphics[width=1\linewidth]{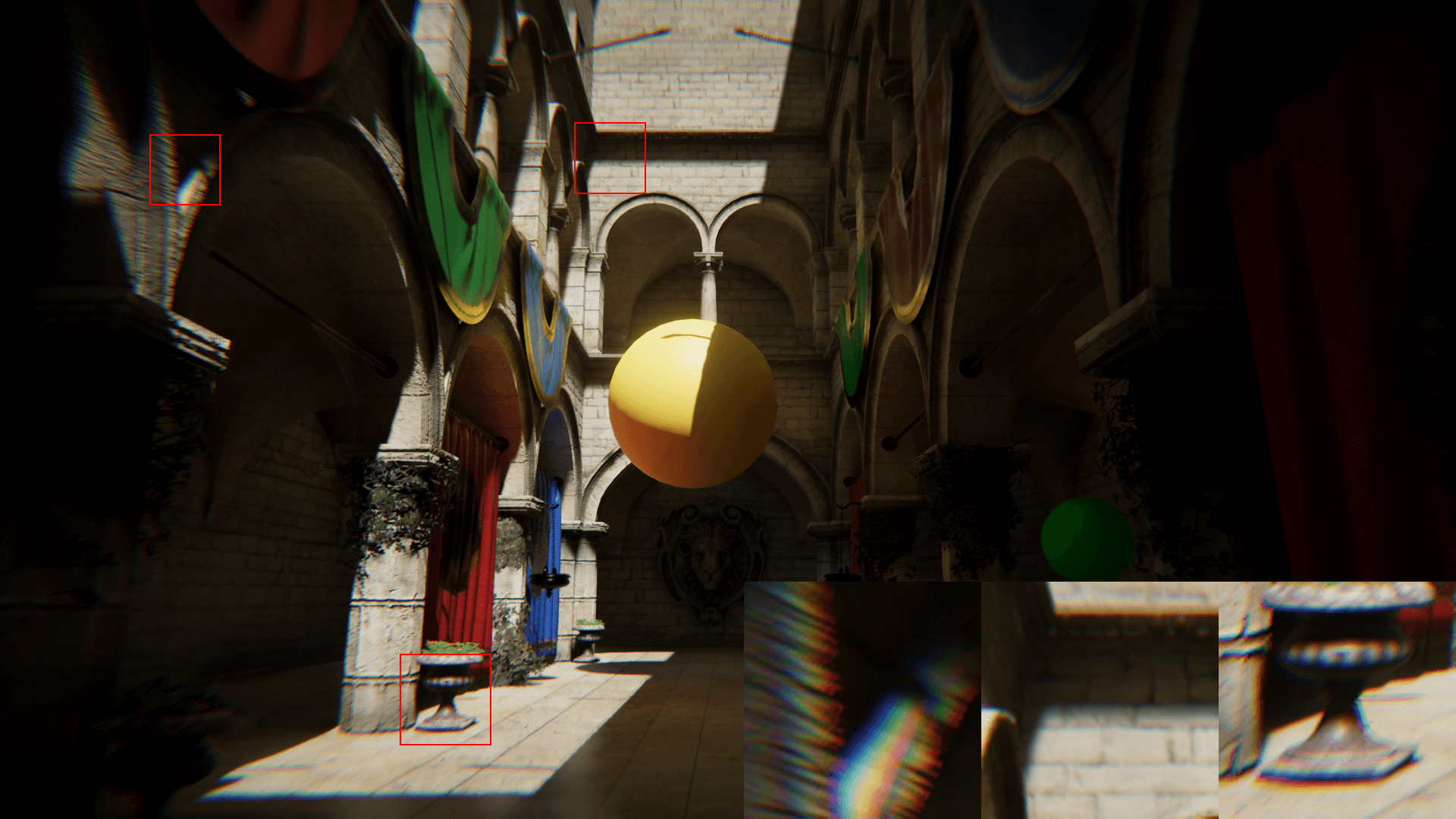}
        \caption{Contact GI off}
    \end{subfigure}

    \caption{Some examples of the effect of Contact GI on a simple lighted room (a) (b) and the Sponza Palace (c) (d). Contact GI greatly compensates the loss of detail produced by the discretization of the irradiance by enhancing the realism of details.}
    \label{fig:contactgi}
\end{figure}

At the same time of computing ambient occlusion, we sample the lighting at occluded points and merge it with the probe GI so that we can take into account the effect of multi-bounce lighting, therefore vastly improving the level of detail of GI.

%-------------------------------------------------------------------------
\subsection{Cascade volume}

By employing a varying-density probe volume, we provide a stratified Cascade Probe Volume, that is, assign a sparsely populated probe volume for further away regions from the camera, making the algorithm feasible for a greater extension of the scene, even more than 1 km away. This way this method can not only be applied for small indoor scenes, but for mid-ranged building scenes and large open worlds as well. Besides, this process of reduction of more distant probes can be adapted to the actual cost of global illumination in order to further extend the effect of global illumination in the scene. We employ Mean Value Coordinates\cite{floater2005mean} instead of trilinear interpolation for the interpolation between different density volumes so that we can have a softer transition.

For the furthest strata of the probe volume we do not perform any visibility test since at this scale it seemed unnecessary for global illumination.

\begin{figure}[h]
    \centering
    \begin{subfigure}{1\linewidth}
        \includegraphics[width=1\linewidth]{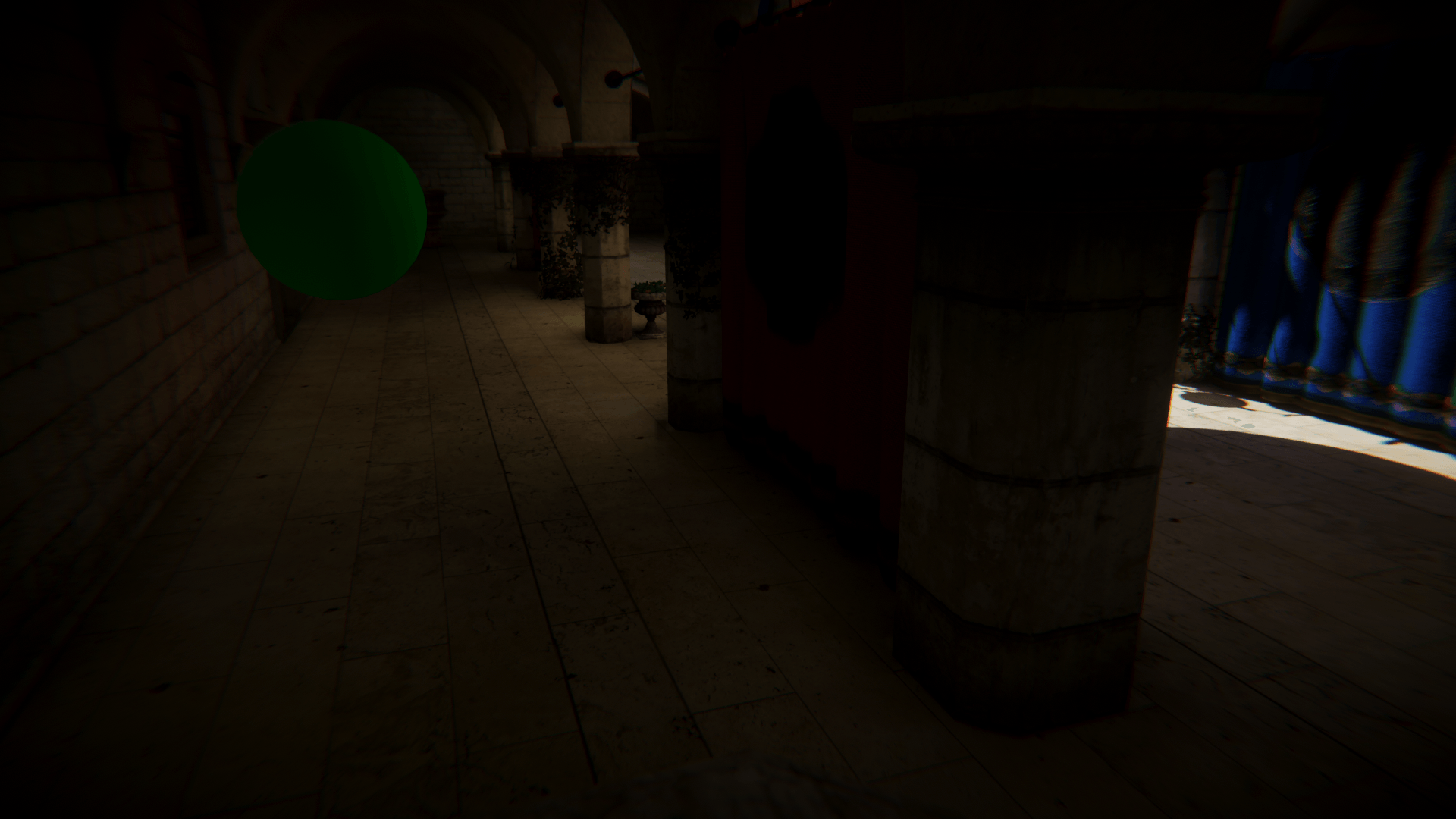}
        \caption{SDFDDGI}
    \end{subfigure}

    \begin{subfigure}{1\linewidth}
        \includegraphics[width=1\linewidth]{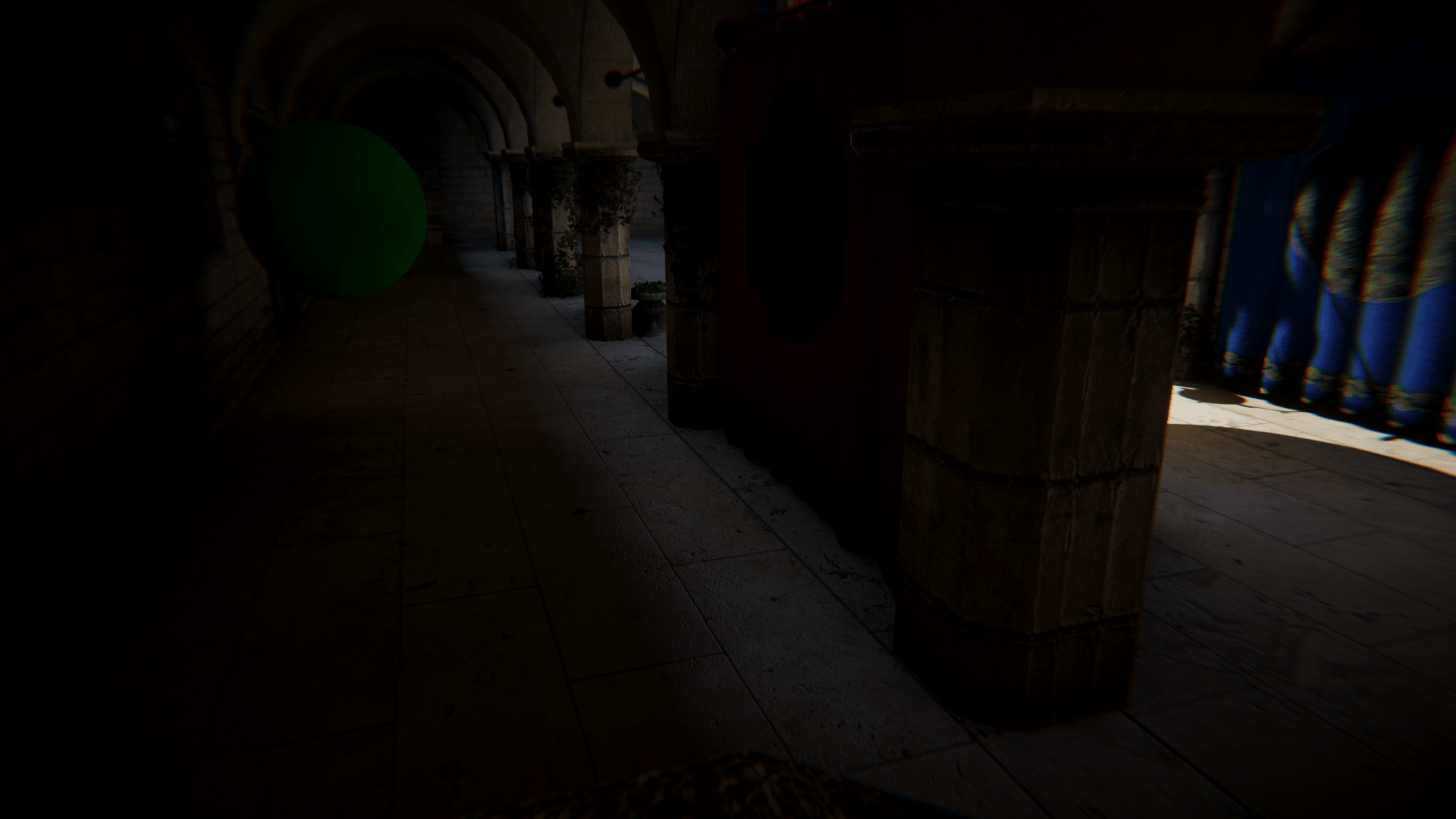}
        \caption{RTXGI}
    \end{subfigure}

    \caption{Comparison of light leaking artifacts on SDFDDGI and RTXGI. Our method (a) is able to eradicate light leaking. However, on thin objects such as the curtain on the Sponza scene RTXGI (b) may face light leaking issues.}
    \label{fig:leakcompare}
\end{figure}

%-------------------------------------------------------------------------
\section{Results}

%我们在不同的场景和硬件平台下测试了我们的方法，测试了它的效果与性能。我们将我们的结果与已有的实时全局光照方法进行了比较。
We tested the performance and the quality of GI of our algorithm with different hardware and scenes. Most of the experiments were performed in comparison to other state-of-the-art real-time global illumination approaches.

%-------------------------------------------------------------------------
\subsection{Probe volume resolution}
%由于probe的位置是在运行时通过SDF梯度下降的方式来分配的，因此完全无需人工干预就能得到理想的probe分布结果。我们首先测试了不同probe volume分辨率对最终结果的影响，并将结果与离线PathTracer得到的GroundTruth做比较。

\begin{figure*}
    \centering
    \begin{subfigure}{0.31\textwidth}
    \caption{2x1}
    \includegraphics[width=1\linewidth]{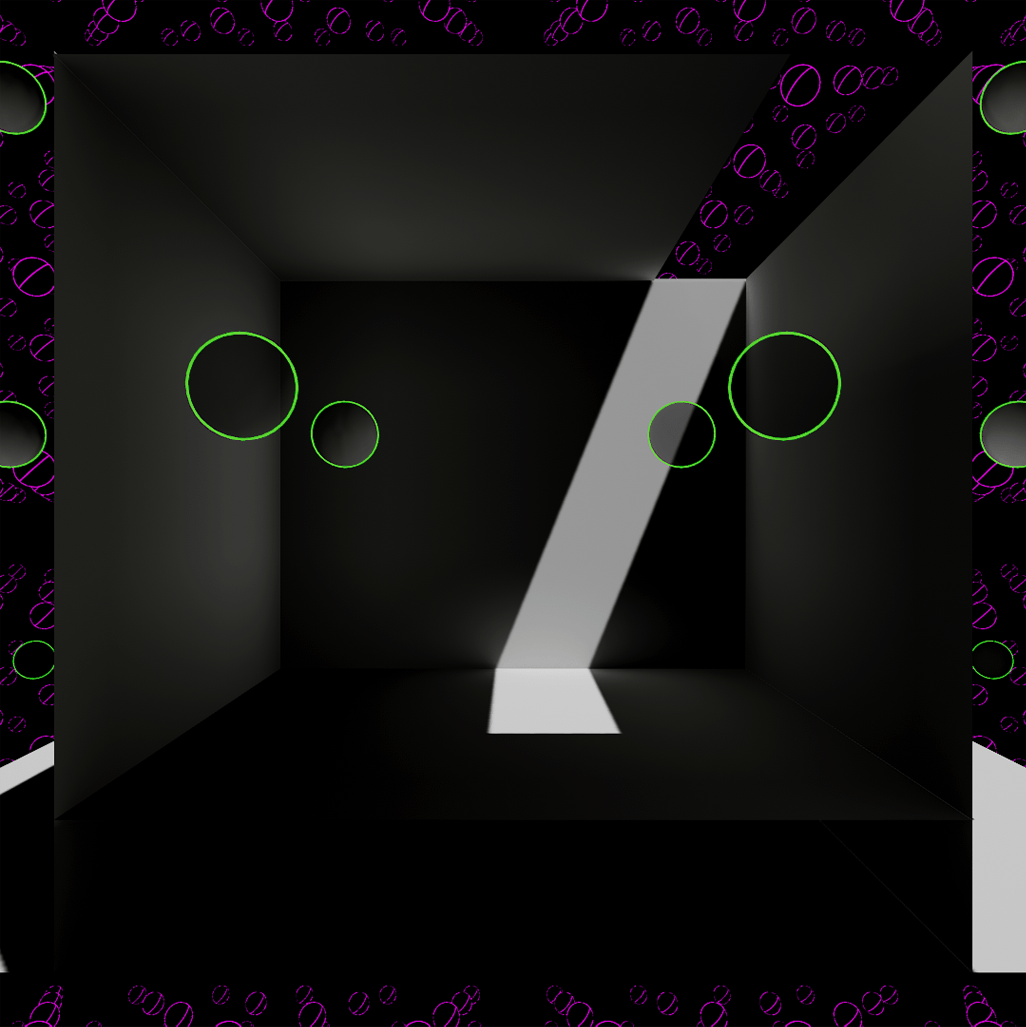}
    \end{subfigure}
    \begin{subfigure}{0.31\textwidth}
    \caption{2x2}
    \includegraphics[width=1\linewidth]{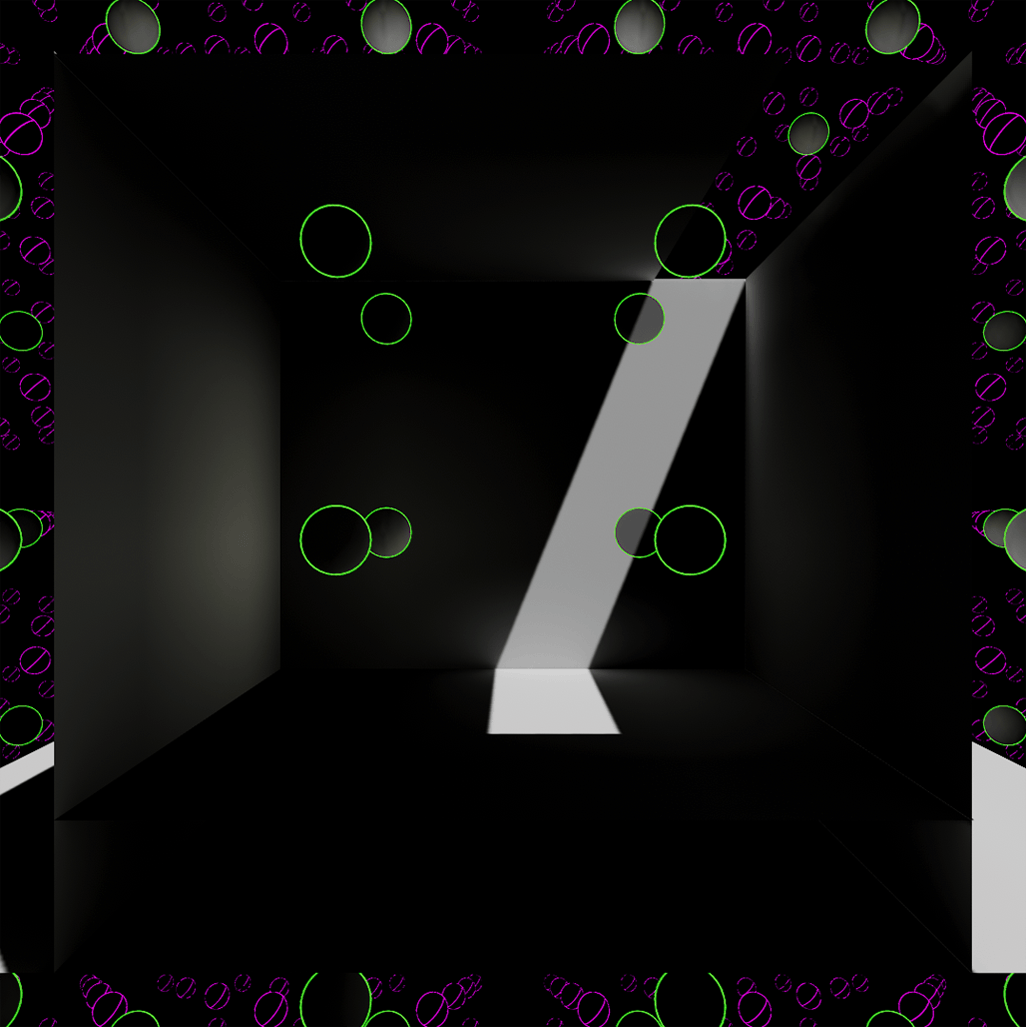}
    \end{subfigure}
    \begin{subfigure}{0.36\textwidth}
    \caption{6x6}
    \includegraphics[width=0.861\linewidth, right]{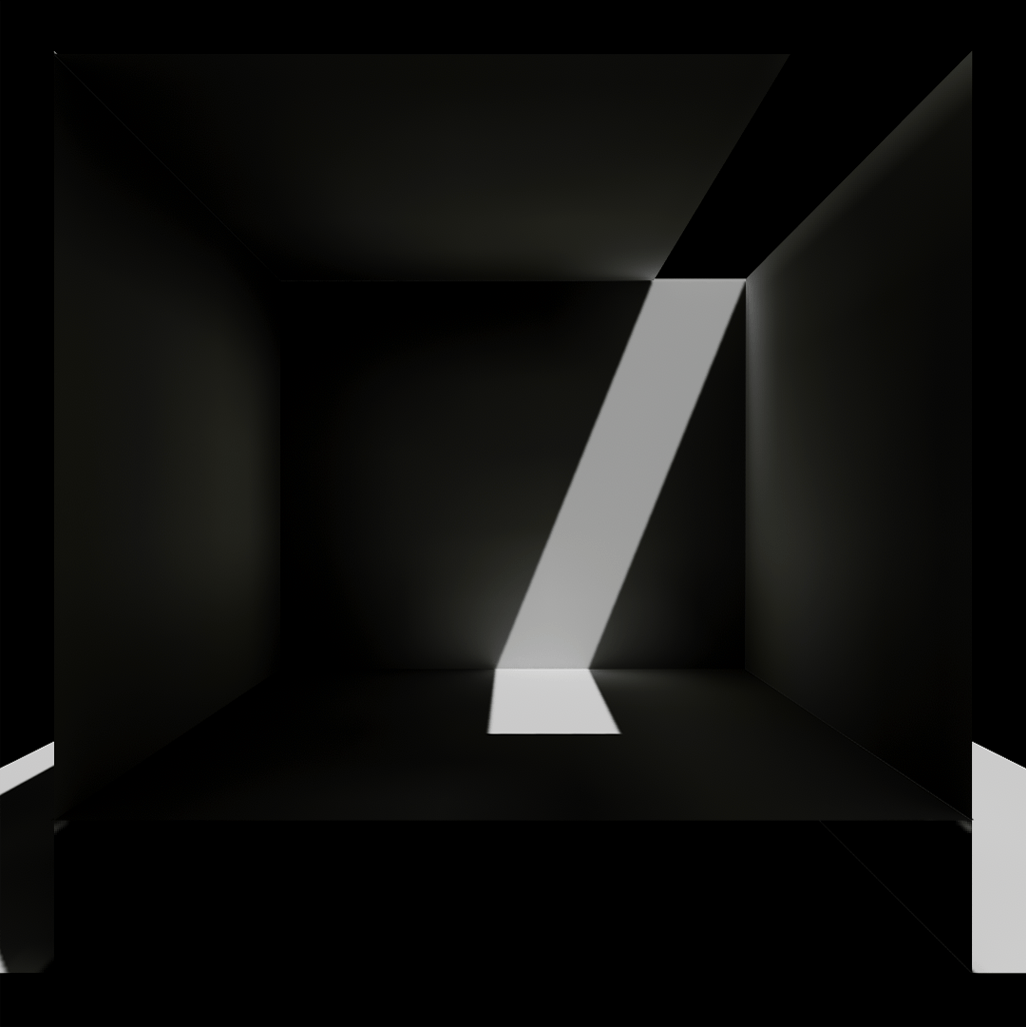}
    \end{subfigure}

    \begin{subfigure}{0.31\textwidth}
    \includegraphics[width=1\linewidth]{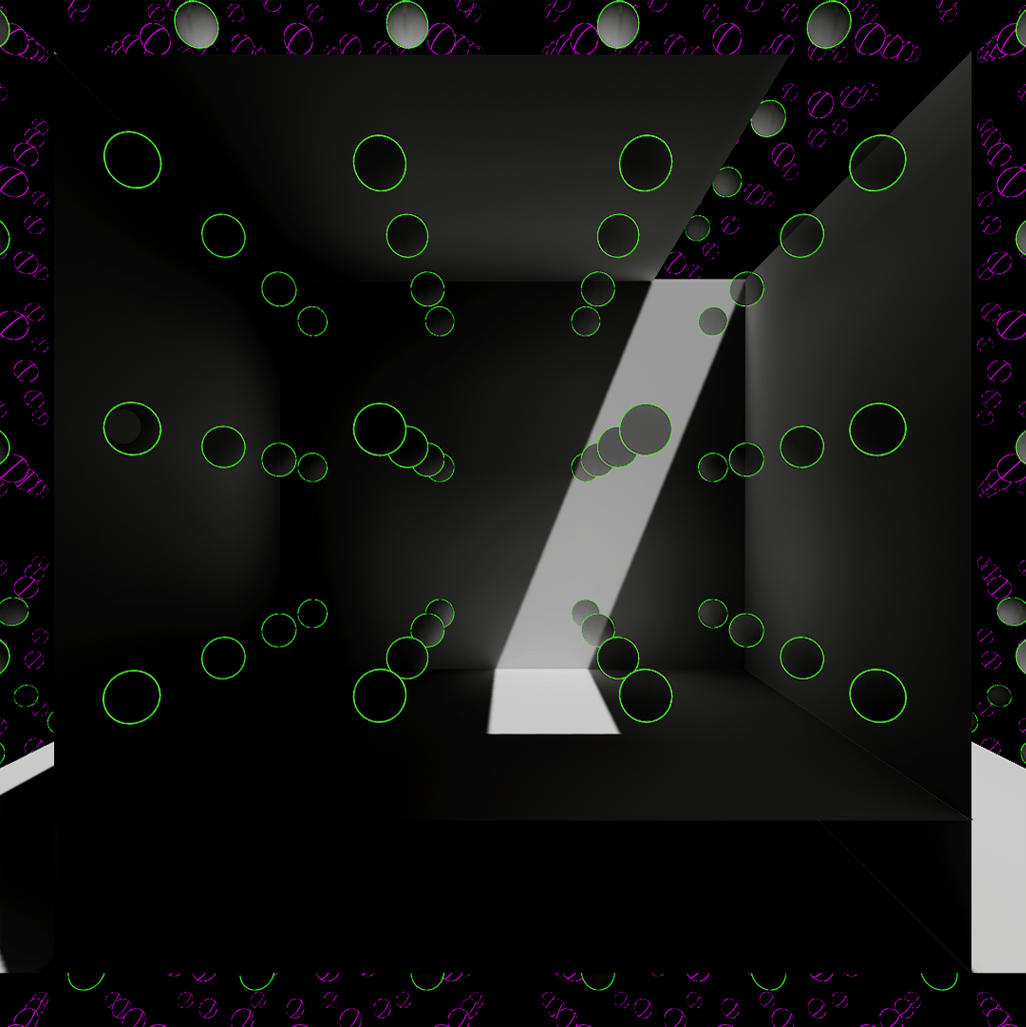}
    \caption{4x4}
    \end{subfigure}
    \begin{subfigure}{0.31\textwidth}
    \includegraphics[width=1\linewidth]{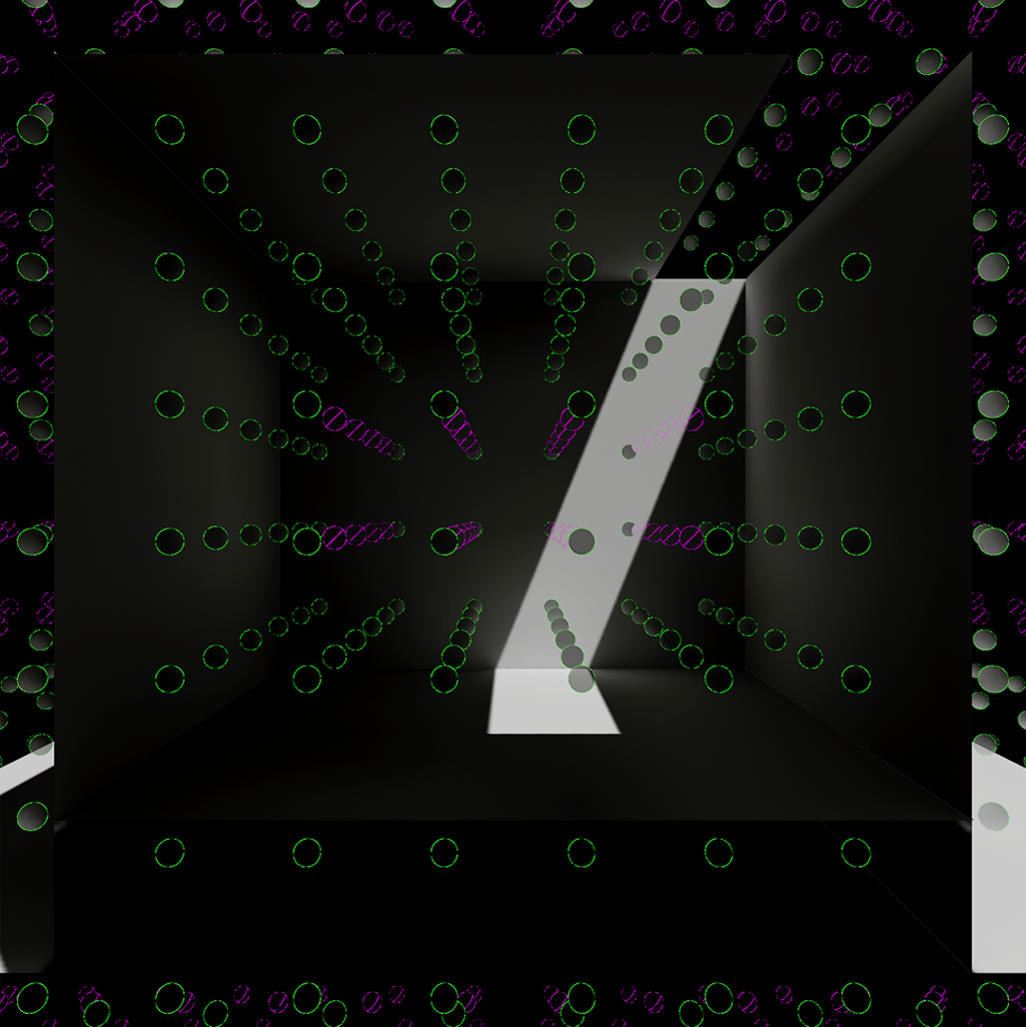}
    \caption{6x6}
    \end{subfigure}
    \begin{subfigure}{0.36\textwidth}
    \includegraphics[width=0.861\linewidth, right]{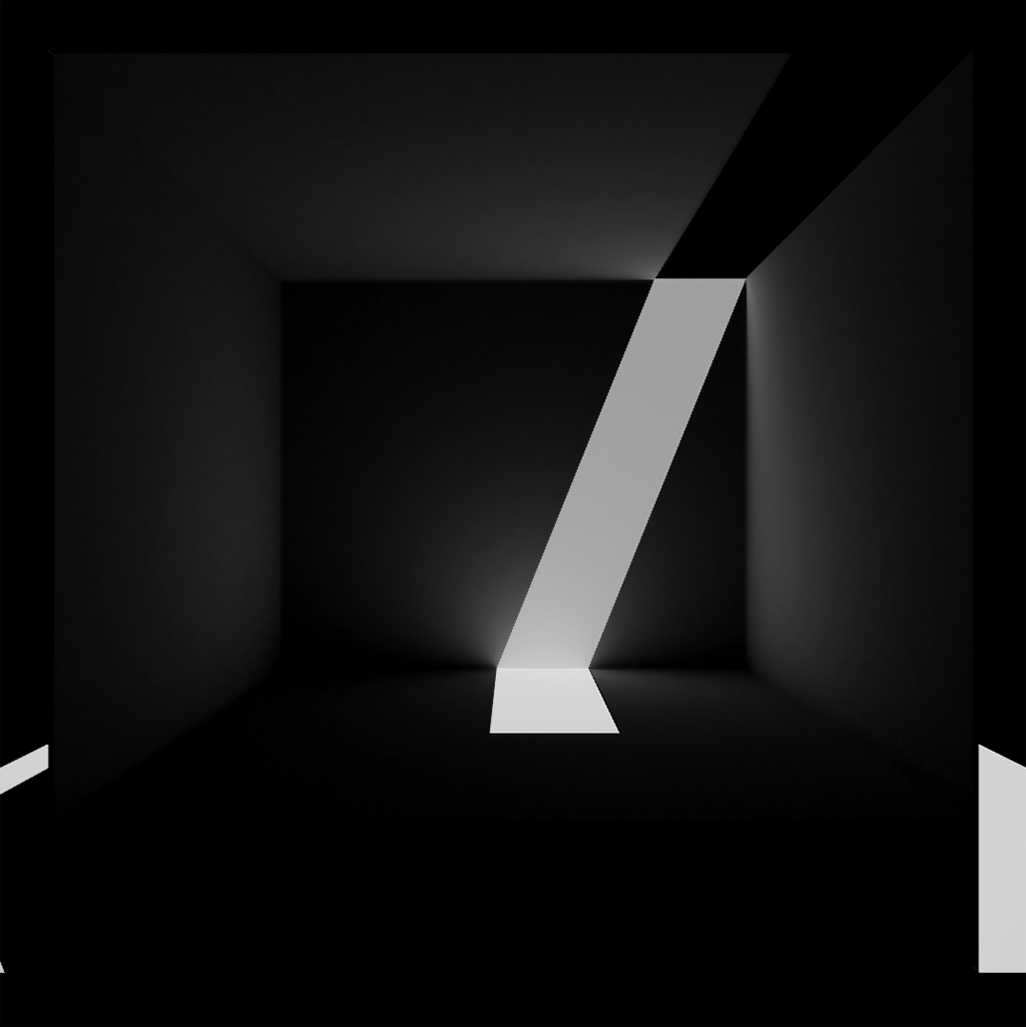}
    \caption{Ground Truth}
    \end{subfigure}

    \caption{SDFDDGI results with different probe volume resolutions. Path Tracing Ground Truth as reference. With too few probes, as we see in the room with 2x1 probes (a), we are not able to produce satisfying global illumination but light leaking is still unpresent. Besides, with a denser, but still relatively rough, probe volume grid (c), we can get similar global illumination as in a Path Tracing reference (f).}
    \label{fig:probedensity}
\end{figure*}

Since the position of the probes is assigned at runtime according to SDF gradient descent, this method does not need any previous human intervention for an ideal distribution of the probes. We first observed the effect of the resolution of the probe volume on the result GI and then we compare it to an offline path tracer as Ground Truth.

In Figure \ref{fig:probedensity} we can observe that if probe volume resolution  is very low it will not produce light leaking problems, although the result GI will, of course, not be very accurate.

%从图中可以看到，我们的方法在probe分辨率极低时，虽然会导致结果发生较大的偏差，但是仍然不会出现漏光问题。

 %-------------------------------------------------------------------------
\subsection{Effect of Contact GI}
 %我们对比了开关contact gi带来的结果差异。
Contact GI is able to compensate the loss of detail of diffuse GI produced by the sampling of the irradiance function. In Figure \ref{fig:contactgi} we can observe the difference of using Contact GI on a basic lighted room and on the Sponza scene.

Contact GI brings about significant improvements to the effect of minor traits of the scene to the diffuse global illumination. It solves some of the loss of detail caused by the disparity of the position of the probes and the real position taken into account for shading.

 %从图中可以看到，Contact GI带来了GI细节质量的显著改善，解决了一部分因为Porbe位置与实际shading位置不符造成的细节丢失问题。

 %-------------------------------------------------------------------------
\subsection{GI comparison to other methods}
 %我们在Sponza场景测试了我们方法，并将结果与其他已有的实时GI技术进行了对比。该场景共有25w个三角面，26种不同材质和48张不同分辨率的纹理。我们使用43个Primitives进行场景的SDF场进行动态构建，并将其划分为8个Cluster。SDFDDGI与RTXGI的Probe Volume分辨率为22x14x32。VXGI的体素分辨率为64x64x64。

We compared our method with other real-time GI techniques using the Sponza scene. This scene has 5,000 triangles, 26 different materials and 48 different resolution textures. We employed 43 SDF primitives for the dynamic representation of the structure of the scene and fall it apart into 8 clusters. The probe volume of SDFDDGI, as well as RTXGI, has a resolution  of 22x14x32 probes. VXGI voxel resolution is 64x64x64.

In Figure \ref{fig:gicompare} we can see that with the same probe volume, our approach fits better the subtleties of the scene. In comparison to Ray Tracing GI, our method is able to support multiple bounce illumination and at the same time it does not need a denoiser, which could blur some details. VXGI with a higher resolution is not able to achieve the same level of detail and, at the same time, it already needs longer computation time.

% 从图中可以看到，在相同的Probe Volume配置下，我们的方法比RTXGI有更精细的细节。相较于RayTracingGI，我们支持多次反射，且不会因为降噪而造成细节的模糊。同样的，VXGI在较高分辨率下，仍然无法达到我们方法的精度，且在该配置下已经需要更多的计算时间。

As seen in Figure \ref{fig:leakcompare}, for low resolution probe volume there is no light leaking, which is still a problem for RTXGI for, for example, thin objects as the curtains. Instead our method is able to correctly manage the occlusion of curtains.

%此外从上图中可以看到，对比RTXGI，我们的方法完全解决了漏光问题，厚度很薄的幕布会造成RTXGI发生leaking，而我们的方法很好的处理了幕布产生的遮挡。

 %-------------------------------------------------------------------------
\subsection{Performance}

% 我们以Sponza宫为例，进行了横向的性能测试。各种技术方案在本节中所使用的配置为前文进行效果对比时使用的配置。

We tested the performance of the algorithm with Sponza as test scene under the same configuration as in previous section, analyzing the time consumed for every stage of the algorithm, also in comparison to the other aforementioned methods.

%表：性能

\begin{table}[h]
    \begin{center} {\footnotesize

    \renewcommand{\arraystretch}{1.5}
    \begin{tabular}{ccc}
    \hline
    % Header
     & \multicolumn{2}{c}{Consumed time}  \\
     & \multicolumn{1}{c}{Total} & \multicolumn{1}{c}{Per stage}\\
    \hline
    % Table
    \multirow{3}{*}
    &&
        \textit{Probe Update}: 0.70\\[-2ex]
        {SDFDDGI}&1.67&\textit{Contact GI}: 0.41\\[-2ex]
        &&\textit{Shading}: 0.56\\

    \multirow{3}{*}
    &&
        \textit{Depth Mipmap}: 0.06\\[-2ex]
        {SSGI (diffuse only)}&1.17&\textit{HiZ trace}: 0.69\\[-2ex]
        &&\textit{Denoise}: 0.42\\

    \multirow{2}{*}
        \ RTXGI&3.98&\textit{Probe Update}: 3.28\\[-2ex]
        &&\textit{Shading}: 0.70\\

    \multirow{2}{*}
        \ Ray Tracing GI (diffuse only)&4.13&\textit{Trace ray}: 2.23\\[-2ex]
        &&\textit{Denoising}: 1.90\\

    \multirow{2}{*}
        \ VXGI (diffuse only)&5.24&\textit{Voxelize}: 3.31\\[-2ex]
        &&\textit{Cone trace}: 1.93\\

    \hline
    \end{tabular} }
    \end{center}
    \caption{\footnotesize Stage by stage performance comparison of state of the art real-time GI approaches with our method, all running on RTX 2080Ti and I7 9700k, with a render resolution of 1920x1080 on the Sponza scene.}
    \label{table:performance}
\end{table}

As we see in Table \ref{table:performance}, our method achieves the best performance for the same or better quality. Time per frame is only shorter for Screen Space GI, which makes a lot of quality compromises in order to achieve this performance.

%如上表所示，我们的方法在实现非常好的质量的同时，在不同硬件平台上都达到了几乎最为优秀的性能，在耗时上仅次于屏幕空间方法。

%-------------------------------------------------------------------------
\subsection{Conclusions}
We have proposed a novel approach to calculate real-time global illumination using SDF. Apart from saving all work flow for baking, and supporting fully dynamic lighting and geometry for a scene, it also provides better quality GI and has higher performance than other similar approaches, thus having some potential applications. However, our method still has room for improvement. For example, according to the relative placement of the camera and the probes, we could use importance sampling around this direction in order to further stabilize global illumination because only the normal facing to camera can be seen by the camera. Our research interest also focuses on dynamic GI, so for specular GI we still have to rely on a mixed approach using other methods such as SSR or Ray Tracing, but using SSR on top does not add any extra cost to achieve diffuse-specular path. Last of all, our approach uses simplified SDF primitives to represent the scene, until now we manually provide its simplified SDF representation, what requires an enormous amount of work for large and complex scenes. For the future, it would be necessary to research on the automatization of this process.

%-------------------------------------------------------------------------
% bibtex
\bibliographystyle{eg-alpha-doi}
\bibliography{paper}

% biblatex with biber
% \printbibliography

\end{document}